\documentclass[a4paper,11pt]{aa}

\usepackage{graphicx}
\usepackage{subcaption}
\usepackage{rotating}
\usepackage[T1]{fontenc}
\usepackage{multirow}
\usepackage[]{algorithm2e}
\usepackage[title]{appendix}
\usepackage{linenoaa}
\usepackage{bm}
\usepackage[normalem]{ulem}
\usepackage{txfonts}
\usepackage{hyperref}
\usepackage{xcolor}

\usepackage{multirow}
\usepackage{comment}
\usepackage{makecell}
\usepackage{enumitem}
\usepackage{dashrule}
\usepackage{xspace}
\usepackage{booktabs}

\newcommand{\fermi}{{\em Fermi}}

\usepackage{comment}

\begin{document}
\title{Luminosity function of millisecond pulsars in globular clusters from their gamma-ray emission}

\author{Francesca Calore\inst{1}
        \and
        Christopher Eckner\inst{2, 3, 4}
        \and
        Silvia Manconi\inst{1,5}
        \and
        Francesco Xotta\inst{4}
        \and
        Gabrijela Zaharijaš\inst{4}}

\institute{
LAPTh, CNRS, F-74000 Annecy, France\\
\email{calore@lapth.cnrs.fr}
\and
Instituto de Astrofísica de Canarias, E-38205 La Laguna, Tenerife, Spain
\and
Universidad de La Laguna,
Dpto.~Astrofísica, E-38206 La Laguna, Tenerife, Spain\\
\email{ceckner@iac.es}
\and
Center for Astrophysics and Cosmology, University of Nova Gorica, Vipavska 13, SI-5000 Nova Gorica, Slovenia\\
\email{francesco.xotta@ung.si, gabrijela.zaharijas@ung.si}
\and
Sorbonne Université \& Laboratoire de Physique Théorique et Hautes Énergies (LPTHE), 4 Place Jussieu, Paris, France\\
\email{manconi@lpthe.jussieu.fr}}

\abstract
{Millisecond pulsars (MSPs) are widely believed to be responsible for the gamma-ray emission of globular clusters (GCs), yet the underlying MSP gamma-ray luminosity function remains uncertain. Existing GC-based determinations of the latter often rely on integrated fluxes and on external prescriptions for the number of MSP in each cluster.}
{We aim to constrain the MSP gamma-ray luminosity function in Milky-Way GCs directly from gamma-ray spectral energy distributions (SEDs), and to establish an inference strategy that remains computationally viable for joint analyses of large GC samples.}
{We construct a forward model in which each GC hosts a population of MSPs whose luminosities are drawn from a universal log-normal luminosity function parameterized by a pivot luminosity $L_{0}$ and width $\sigma_{L}$. Synthetic GC SEDs are generated by summing individual MSP spectra, and we infer posterior distributions using (i) likelihood-based nested sampling and (ii) { Simulation Based Inference (SBI)} in the neural ratio estimation (NRE) approach. The SBI posteriors are validated on synthetic datasets through comparisons with the corresponding likelihood-based results and through simulation-based coverage tests. We then apply the SBI pipeline to \textit{Fermi}-LAT SEDs from the 4FGL-DR4 catalog for a baseline sample of 36 GC-associated sources, augmented by dedicated phase-resolved ON/OFF SEDs for two clusters hosting individually detected MSPs.}
{The likelihood-based nested sampling approach becomes rapidly impractical as the number of jointly modelled clusters (and cluster-specific latent parameters) increases, whereas the NRE approach scales mildly with sample size after amortized training and reproduces the likelihood-based posteriors in regimes where direct sampling is feasible. For the baseline real-data sample, we obtain informative constraints on the luminosity-function parameters, $\log_{10} L_{0} = 33.3^{+0.2}_{-0.3},\;
\sigma_{L} = 1.2^{+0.3}_{-0.5}$ (68\% credible intervals) consistent with previous analyses that constrain the cluster-specific MSP counts via complementary multi-wavelength information. Novel phase-resolved ON/OFF SEDs for two GC are derived. Together with clusters dominated by a single bright MSPs, adding these information is shown to add critical constraints to the luminosity function parameters.
}{}


\titlerunning{MSPs luminosities from GC gamma-ray spectra}
\maketitle
\nolinenumbers

\section{Introduction}
\label{sec:intro}

Globular clusters (GCs) are dense, gravitationally bound systems of stars, typically found in the halos of galaxies. They are among the oldest objects in the Universe, hosting stellar populations that provide unique laboratories for studying stellar evolution, dynamics, and exotic astrophysical phenomena~\citep{2019A&ARv..27....8G}. GCs are identified through their compact, spherical morphology and distinct stellar content, which includes a mix of low-mass stars, blue stragglers, and millisecond pulsars (MSPs). Their high stellar densities promote frequent dynamical interactions, making them prolific factories for the formation of compact objects, including MSPs~\citep{2025arXiv250814308K}, as also confirmed by modern N-body simulations~\citep{2019ApJ...877..122Y}.

MSPs in GCs are of particular interest as probes of both stellar dynamics and high-energy astrophysics. Their gamma-ray emission, detected by the \textit{Fermi} Large Area Telescope (\textit{Fermi}-LAT, see \cite{2009ApJ...697.1071A}), offered a window into the population properties of MSPs and the physical processes governing their formation and evolution~\citep{2010A&A...524A..75A} \cite{Gonthier:2018ymi,2021MNRAS.507.5161S,deMenezes:2023mzx,Henry:2024mdn,Amerio:2024qor}. The gamma-ray emission from GCs is predominantly interpreted as the cumulative contribution of a population of MSPs~\citep{1991Natur.352..695C}, a hypothesis supported by the gamma-ray spectral behaviour and the fact that pulsation is detected from four objects (NGC 6624, NGC
6626, NGC 6652 and NGC 6341) \cite{Zhang:2023gwr}. These four objects, which we will refer to as \emph{special GCs} in what follows, were analysed while accounting for the periods during which the pulsars detected within them exhibited emission beamed towards the observer (see \cite{10.1093/mnras/267.1.125, Lyne_1987, Gautam_2022, Pan_2020}). We refer to these periods as ``ON-phase'', whereas we refer to the other periods as ``OFF-phase'' throughout this work. Moreover, NGC 6624 and NGC 6626 exhibit significant gamma-ray emission when selected for being ON-phase with the gamma-ray pulsation of their respective MSPs as well as when being OFF-phase~\cite{Zhang:2023gwr}, indicating the existence of at least one additional MSP per GC.
 However, the precise nature of a GC's gamma-ray emission, including the presence of a possible high-energy tail, and the justification for attributing it to MSPs remain active areas of investigation~\citep{2021MNRAS.507.5161S}.
 { In this study, we work under the assumption that all the GC's gamma-ray emission can be explained by the cumulative emission of the MSPs contained in it.} 

MSPs are well-known gamma-ray emitters, typically modeled, as in the Third \textit{Fermi}-LAT Pulsar Catalog (3PC)~\citep{Fermi-LAT:2023zzt}, through a power law with exponential cutoff spectrum. Studies such as \cite{2024MNRAS.530.3552L} have characterized the cumulative gamma-ray spectrum of MSP populations, but significant uncertainties persist, particularly regarding the gamma-ray luminosity function (GLF). 
The GLF describes the distribution of intrinsic luminosities among MSPs and is a critical observable for understanding their contribution to the gamma-ray sky. Determining the GLF is essential also for leveraging GCs as probes of the broader MSP population in the Galaxy.

Several works have inferred the gamma-ray luminosity function of MSPs in the Galactic disc (also referred to as ``field MSPs'' below), as well as in GCs using \textit{Fermi}-LAT source catalogs and complementary, multi-wavelength proxies. 

In the case of field MSPs, this is typically done by forward-modeling a Galactic spatial distribution together with a luminosity function and an instrument-dependent selection function. 
\citet{Hooper:2015jlu} use the 3FGL catalog {(based on 4 years of {\it Fermi}-LAT data, see \cite{latcollaboration2015fermilargeareatelescope})} and fit the angular and flux distributions of pulsar-labeled sources under an assumed disc model for their spatial distribution, effectively avoiding the use of individual distance estimates for MSPs. In contrast, \citet{Bartels:2018xom} perform a Bayesian population analysis that incorporates available distance information and marginalizes over the associated uncertainties, thereby reducing the flux-luminosity degeneracy for the subset of sources with informative distance constraints and providing a more complete propagation of distance-related uncertainties into the luminosity function posterior. \citet{Ploeg:2017vai} additionally consider a bulge component motivated by the interpretation of the Galactic Center Excess, which can broaden the allowed luminosity function region through degeneracies between the bulge contribution and the per-source luminosity distribution. More recently, \citet{Holst:2024fvb} updated the field-MSP analysis of \cite{Hooper:2015jlu} (again, without additional distance information) using the 3PC catalogue, which provides a substantially enlarged MSP sample and revised source associations.

Constraints derived from GCs in the present-day literature are typically based on the integrated gamma-ray luminosities of a sample of clusters and an external prescription linking each cluster to an expected MSP content. \citet{Hooper:2016rap} and the updated analysis of \citet{Amerio:2024qor} follow this approach, modeling the cluster luminosities assuming a universal luminosity function while using encounter-rate and/or optical-luminosity proxies to inform the expected MSP number per cluster (including variants with tunable scaling relations). These analyses are conceptually complementary to field-MSP studies: they do not rely on a Galactic spatial model for resolved sources, but they do introduce an explicit dependence on multi-wavelength proxies and on the assumed form and scatter of the MSP-number-to-proxy relation. 

Our goal is to use the gamma-ray properties of detected GCs and, when available, their {phase}-resolved SEDs to infer the GLF of MSPs, which are the most likely sources of this emission.  Unlike previous studies, we aim to derive the GLF from gamma-ray data alone, without external prescriptions based on multi-wavelength proxies.
These prescriptions have been used to break the degeneracy between GLF parameters and the total number of MSPs in each GC. 
Our primary observable is the spectral energy distribution (SED) of each GC, {however, we show in our work how employing the integrated fluxes of GCs may also be a valid choice that yields very similar results to the SED benchmark observable (see Appendix \ref{app:SED-vs-flux}).} { To disentangle the degeneracy between the parameters considered,} we leverage for the first time {phase}-resolved SEDs to enhance our constraints. This approach allows us to probe the GLF in a more model-independent manner, focusing solely on the gamma-ray emission characteristics.

Our methodological novelty lies in the use (and comparison) of two Bayesian strategies: a traditional likelihood-based analysis using nested sampling, and an implicit-likelihood method that replaces the explicit likelihood with an implicitly learned likelihood via simulated samples. For the remainder of this study, we will use the term simulation-based inference (SBI) to refer to this implicit-likelihood approach. While nested sampling provides a robust reference for small target sets, its computational cost scales poorly with the number of clusters. SBI, in contrast, shifts the dominant cost to data simulation and neural network training, enabling scalable posterior inference for large GC samples. 
In recent years, SBI has gained a lot of popularity in fundamental physics, including particle physics, cosmology and astrophysics with applications to both synthetic and real observations\footnote{A track record of SBI works is provided and maintained by S.~Mishra-Sharma at: \href{https://github.com/smsharma/awesome-neural-sbi?tab=readme-ov-file}{Awesome Neural SBI}}. An example of an application of Bayesian SBI to MSPs in globular clusters is reported in \cite{Berteaud:2024zda}, focusing on observed radio MSPs in Terzan 5. Our study aims to add an analysis on (high-level) \textit{Fermi}-LAT spectral data of a large set of  GCs and to demonstrate the superiority of Bayesian SBI compared to nested sampling in this context. 

The structure of the paper is as follows: 
In Sec.~\ref{sec:msp-general}, we describe the modeling of the gamma-ray emission in GCs, the sample selection, and the main (synthetic and real) datasets used in the analysis. In Sec.~\ref{sec:inference} we detail our likelihood-based Bayesian inference framework and SBI. Sec.~\ref{sec:sims-results} presents our tests on sythetic data and validation of the inference SBI framework. 
In Sec.~\ref{results:sbi-LATdata}, we discuss the application of the SBI pipeline to real \fermi-LAT data, inferred GLF and comparisons with previous studies. We discuss the implications of our findings in Sec.~\ref{sec:discussion}, and conclude in Sec.~\ref{sec:conclusion}.


\section{Gamma-ray emission of globular clusters}
\label{sec:msp-general}

\subsection{Luminosity function model}
\label{sec:lum-fct-intro}
The ultimate goal of this work is to develop a pipeline for inferring the GLF of MSPs from the spectral energy distributions of GCs.
We assume that MSPs in all GCs share the same underlying GLF $\mathrm{d}N/\mathrm{d}L_{\gamma}$, which we model using a log-normal distribution:
\begin{equation}
\label{eq:luminosity_function}
    \frac{\mathrm{d}N}{\mathrm{d}L_{\gamma}}=\frac{1}{L_{\gamma}\sigma_L\sqrt{2\pi}}\exp\!{\left(-\frac{(\ln{L_{\gamma}-\ln{L_0}})^2}{2\sigma_L^2}\right)} \, \mathrm{,}
\end{equation}
with free parameters $L_0$, representing the pivot luminosity of the population(s), and $\sigma_L$ denoting the width of the log-normal profile. This is the same parametrization also used in \cite{Amerio:2024qor} and \cite{Hooper:2016rap} targeting MSPs in GCs, but it was also examined in works on field MSPs like~\cite{Bartels:2018xom}.  
{The parameters of the GLF are however not sufficient to determine the gamma-ray emission in a GC, which 
also depends on the total number of MSPs in the cluster, $N_{\rm I}$ (where I indicates the GC name).
This number is highly uncertain, both from a theoretical and an observational point of view (see \cite{Kalambay:2026xqw}). } 
{In the following, we do not rely on any external, multi-wavelength information to constrain $N_{\rm I}$ and we will infer it as an additional free parameter.}

\subsection{Globular clusters sample}
\label{sec:gc-samples}

For our study, we define several samples of GCs that we either use for analysis and parameter estimation or to verify the functionality and robustness of our analysis pipeline and inference methods. 
The exact definitions for each set are stated in Tab.~\ref{tab:gc-sets} of Appendix \ref{app:gc-sets}. We will use these four target datasets to show the impact of the number of considered GCs on the MSP GLF parameter estimation. Additionally, a synthetic GC sample is used to first benchmark the performance of our adopted analysis techniques with simulations.


\paragraph{(I) Synthetic GC sample.} We consider a total of nine mock GCs (see Table \ref{tab:synthetic-SEDs}). The first {six} were chosen due to their intermediate brightness (M 62, NGC 6752, $\omega$-Centauri, NGC 1904, Terzan 2, NGC 6218), as detailed in the next paragraph. We then added NGC 6341, NGC 6652, (both known to host 1 MSP with no significant OFF-phase SED) and NGC 6624 {(as before, but with a significant OFF-phase emission detected. See Sec.~\ref{sec:gc-sed}).}

\paragraph{(II) 4FGL-DR4 GC target samples for real-data analysis.} The most important GC sample selection concerns gamma-ray-bright GCs observed by the \textit{Fermi}-LAT. The authors of \cite{Amerio:2024qor} compiled a list of 56 gamma-ray-bright GCs detected in the 0.1--100 GeV band. We utilize this list as a starting point to compile our own sets of GCs. To this end, we adopt the gamma-ray properties of GCs as stated in 4FGL-DR4~\citep{2022ApJS..260...53A, ballet2024fermilargeareatelescope} by enforcing two main criteria: An eligible 4FGL source must, \emph{(i)}, feature a firm association with a known GC in the form of an association name and, \emph{(ii)}, the distance to the GC must be listed in \cite{2021MNRAS.505.5957B} to ensure a relatively precise (up to 10\%) knowledge thereof.  In addition, we want to ascertain that the derived posteriors for the model parameters have converged given the adopted number of GCs. Therefore, we define a total of four different target datasets with an increasing number of GCs to test convergence (see Fig.~\ref{fig:SBI-real-L0sL} for the final result). We stress that the four special GCs with individually detected MSPs (NGC 6624, NGC 6626, NGC 6652 and NGC 6341) are contained in all the target datasets we prepare.
\begin{itemize}
 \item \textit{Bright target dataset:} It contains a total of eleven GCs comprising the seven brightest GCs in the sky passing conditions \emph{(i)} and \emph{(ii)} as well as the aforementioned four GCs with detected MSPs.
 \item \textit{Intermediate target dataset:} More admissible GCs are added to the bright target dataset in order of decreasing total gamma-ray flux from 0.1--100 GeV to reach a total number of 25 objects.
 \item \textit{Baseline target data:} This target dataset of 36 GCs represents the maximal number of objects passing the conditions \emph{(i)} and \emph{(ii)}. It serves as our benchmark to quote results on real LAT data.
 \item \textit{Extended target data:} Additionally, we include further GCs to our baseline target data by relaxing criterion \emph{(i)}. Some of the GCs of \cite{Amerio:2024qor} are indeed detected as gamma-ray sources in 4FGL-DR4 but lack the association label. We manually associated them if they are within $0.5^{\circ}$ of the GC's optical position (as listed in the \textsc{simbad} astronomical database \cite{2000A&AS..143....9W}). This amounts to 44 GCs. 
\end{itemize}



\subsection{Gamma-ray spectra}
\label{sec:gamma-SED}
As stressed above, the only observable used in this work is the gamma-ray SED of GCs.
In this section, we define what SEDs data products we use when available in the 4FGL-DR4 catalog and illustrate the procedure used to extract the gamma-ray spectra of some of the GCs, when necessary.
We then describe how the synthetic GC SEDs used to test our inference methods on mock data are obtained.

\subsubsection{Synthetic spectra}
\label{sec:generate-synthetic}
We will gauge and validate the performance of our statistical analysis approaches on synthetic gamma-ray SEDs of GCs. Simulations permit us to have full control over the injected parameters. We can also ensure that the generated mock data implement our assumption that MSPs in GCs share the same underlying gamma-ray luminosity function $\mathrm{d}N/\mathrm{d}L_{\gamma}$ represented by a log-normal distribution in Eq.\ref{eq:luminosity_function}.


{Since our target observables are the SEDs of GCs, we generate synthetic SEDs directly. We do not simulate photon events or mock counts cubes, nor do we perform a \texttt{fermipy}~\citep{2017ICRC...35..824W} analysis on simulated data (such a pipeline is applied exclusively to the observational data in Section~\ref{sec:gc-sed} and Appendix~\ref{sec:fermi-data}).}

Yet, having only SED data points is not sufficient for the analysis, because we also need the associated observational uncertainties. To obtain these uncertainties, we start from the observed 4FGL-DR4 SEDs of GCs. For each cluster, we simulate an MSP population whose total SED matches that of the real data from {\it Fermi}-LAT closely enough so that we can adopt the measured uncertainties as provided in the 4FGL-DR4 without generating mock photon counts for the GCs. 

To simplify the process, we assume that all simulated MSPs follow a Power Law with Exponential Cutoff 4 (PLEC4) \footnote{{See \href{https://fermi.gsfc.nasa.gov/ssc/data/analysis/scitools/source_models.html\#PLSuperExpCutoff4}{the Science Tools models}}} spectrum (see Eq.~\ref{eq:PLEC4})  with {representative} spectral parameters: $ E_0  =1.21$ GeV, $ \Gamma  = 1.91$, $ d  = 0.55$, $ b  = 0.73$. 

We fix $L_0^s=3\times10^{33}$ erg/s and $\sigma_L^s=1.6$ for the generation of the synthetic MSP populations. The distances $d_I$ to each respective GC are adopted from \cite{2021MNRAS.505.5957B}. 

For each GC, we simulate 250 synthetic populations for an injected total number of MSPs $N^s_I$, where $I$ is a placeholder for the GC's name. For each individual MSP, we draw luminosities, ${L_c}$ ($c=1,...,N^s_I$), from the fixed luminosity function. Then, we determine the normalization $k_c$ for each MSP by requiring that its flux satisfies the following equation:
\begin{equation}
\label{eq:msp-norm}
F_{\rm{MSP},c} \equiv \frac{L_c}{4\pi {d_{I}}^2}=\int_{0.1\rm\; GeV} ^{100\;\rm GeV}k_c\times \frac{dN}{dE}(E)\times E \, dE.   
\end{equation}
This flux is expressed in units of erg/s/cm$^2$. Lastly, we express the cumulative flux of all MSPs in the $k$-th energy bin (by adopting the energy bin definitions of the reported 4FGL-DR4 source SEDs) as: 
\begin{equation}
\label{eq:msp-cumul}
    F_{\rm MSP}^k =\sum_{c=1}^{N_I^s} \int_{E_{{\rm min},k}}^{E_{{\rm max},k}} k_c\times \frac{dN}{dE}\Bigg|_c(E)\times E \, dE  \rm{.}
\end{equation}
{ where $E_{{\rm min},k}$ and $E_{{\rm max},k}$ represent the minimum and maximum energy at the k-th bin.}
{We checked and verified that a GC SED can be well fitted by a PLEC4 characterized by the representative parameters stated above (see Appendix \ref{app:synthetic-preparation}) and, thus, we proceed to} fix $N^s_I$ to the value that creates an SED (the collection of all $F^k_{\rm MSP}$) whose 68\% uncertainty encompasses the 4FGL-DR4 uncertainty for {all the provided spectral points above 100 MeV. In case of upper limits in the 4FGL-DR4 data points, the synthetic spectrum must not exceed }{ the 68\% upper limit.}

The thereby derived size of the respective GC's MSP population is reported in Tab.~\ref{tab:synthetic-SEDs} of Appendix \ref{app:gc-sets}, and we illustrate via the example of M 62 how we match a synthetic MSP population with the measured gamma-ray emission in Fig.~\ref{fig:synthetic-sed-example} of Appendix \ref{app:synthetic-preparation}.

{Since we want to investigate what the added value of including GCs with detected gamma-ray pulsations is, we also simulate three GCs with this property.}
In particular, for simulations of NGC 6341 and NGC 6652, which were reported to host only one known MSP without significant OFF-phase emission~\citep{Zhang:2023gwr}), we set $N^s_{\mathrm{NGC6341}}=N^s_{\mathrm{NGC6652}}=1$. 
In contrast, NGC 6624 features also significant OFF-phase gamma-ray emission, implying an MSP population larger than 1. 
To account for this, we generate two synthetic SEDs for NGC 6624: A first one where a single bright MSP matches the SED reported in 4FGL-DR4 and a second one at the level of NGC 6218's 4FGL-DR4 SED generated by a certain number $N^s_{\mathrm{NGC6624,OFF}}$ of dim MSPs (drawn from the same MSP luminosity function). We select NGC 6218's gamma-ray SED as a proxy for the OFF-phase emission of our synthetic NGC 6624 because it is an example of a rather faint GC whose integrated gamma-ray flux is only about 20\% of the integrated emission of NGC 6624 \citep{Amerio:2024qor}.  The sizes of each MSP population are also reported in Tab.~\ref{tab:synthetic-SEDs}.

\subsubsection{Real-data spectra}
\label{sec:gc-sed}
{For the 36 GCs of the \textit{baseline target data sample}  we adopt the SEDs as reported in the \texttt{Flux\_Band} category of 4FGL-DR4. We adopt phase-average spectra for NGC 6341 and NGC 6652 despite them hosting an individual MSP.
}

However, we also analyze \textit{Fermi}-LAT {photon events} for some GCs when necessary to complete the SED products {(see Appendix~\ref{sec:fermi-data} for the description of the data selection)}.
More specifically, we conduct a phase-resolved analysis of NGC 6624 and NGC 6626, in which 
individual MSPs have been detected through pulsation identification as well as a significant relic OFF-phase emission~\citep{Zhang:2023gwr}.
Due to their spectral information being unavailable in past literature, we analyze their emission ourselves, separating intervals when the pulsar is ON-phase and when it is OFF-phase.

We assign a pulse phase (a specific identifiable point within the pulsar's full rotational cycle defined between 0 and 1) to each event using the \texttt{tempo2} package \cite{tempo2} together with the \texttt{fermi} plugin \cite{fermiplugin}. The ephemerides for the MSPs are provided by the Fermi Science Support Center\footnote{\url{https://fermi.gsfc.nasa.gov/ssc/data/access/lat/ephems/}}. By examining the resulting phaseograms (constructed by first computing the weight of each event using the \texttt{gtsrcprob} function from the Fermi Science Tools version 1.4.0\footnote{See the \href{https://fermi.gsfc.nasa.gov/ssc/data/analysis/documentation/}{FermiTools Documentation}}), we identify the ON-phase and OFF-phase intervals. 
The ON- and OFF-phases periods are included in Table \ref{tab:onofftable} of Appendix \ref{app:sed-on-off}.

\begin{table}[]
\renewcommand{\arraystretch}{1.6}
    \centering
    \begin{tabular}{lcc}
    \hline
      Name   & $\phi_{\rm on}$ & $\phi_{\rm off}$  \\
      \hline
      NGC 6624 (PSR J1823-3021A)   & 0.62 -- 1.15 & 0.15 -- 0.62 \\
      NGC 6626 (PSR B1821-24) & 0.94 -- 1.59 & 0.59 -- 0.94 \\
    \end{tabular}
    \caption{ON- and OFF-phase intervals for GCs with associated pulsar emissions for which we derive the SED.}
    \label{tab:onofftable}
\end{table}
To obtain their individual ON-phase and OFF-phase spectral energy distributions (SEDs), we develop an analysis chain using \texttt{fermipy v1.4.0}, modifying the configuration file generated by \textsc{EasyFermi v2.0.20}\footnote{See \href{https://github.com/ranieremenezes/easyfermi}{ EasyFermi's Github Page}}  \citep{EasyFermi}. We select \textit{Fermi}-LAT data around the locations of these GCs as described in Appendix \ref{sec:fermi-data} and create a model of all emission sources in the Region of Interest (RoI). The model components are 
the isotropic diffuse model corresponding to the chosen event selection \texttt{iso\_P8R3\_SOURCE\_V3\_v1}, the Galactic diffuse emission model \texttt{gll\_iem\_v07}\footnote{See \href{https://fermi.gsfc.nasa.gov/ssc/data/access/lat/BackgroundModels.html}{{\it Fermi}-LAT Background Models} and \href{https://fermi.gsfc.nasa.gov/ssc/data/analysis/software/aux/4fgl/Galactic_Diffuse_Emission_Model_for_the_4FGL_Catalog_Analysis.pdf}{\texttt{gll\_iem\_v07} Development Report} }~\citep{2016ApJS..223...26A}, and gamma-ray sources in the RoI from the 14-year 4FGL-DR4 catalog (\texttt{gll\_psc\_v35}, see\cite{ballet2024fermilargeareatelescope}). Following \cite{ballet2024fermilargeareatelescope}, we adopt for this study likelihood weights. { Our weights are, however, not created from data, but} {from} models here generated.

After preparation of the dataset, we remove the 4FGL-DR4 source associated with the GC of interest and replace it with two newly defined sources. Since we expect the GC gamma-ray emission to arise from MSPs, we thus assume the functional form which is found to best fit the individual MSPs' spectra collected in the 3PC catalog  to describe their cumulative energy spectrum. Specifically, we model the GC ON-phase spectra as a PLEC4  of the form \cite{Fermi-LAT:2023zzt}:
    \begin{equation}
    \label{eq:PLEC4}
        \frac{\mathrm{d}N}{\mathrm{d}E}=N_0\left(\frac{E}{E_0}\right)^{-\Gamma+\frac{d}{b}}\exp\!{\left[\frac{d}{b^2}\left(1-\left(\frac{E}{E_0}\right)^b\right)\right]} .
    \end{equation}
    where the parameters $\Gamma, d, b, N_0$ are left free, and $E_0=1$ GeV. Since we do not expect to have enough statistical information to constrain a possible cutoff in the spectrum, for the OFF-phase we model the emission as a Power-Law instead. 
As for the background models, we modulate the spectrum of the galactic diffuse emission with a broken power law (with the break energy set to 500 MeV) and limit the normalization of the isotropic diffuse model between 0.8 and 1.2 . {This is done to minimize residuals at energies below $\sim500$} {MeV}.\\
Upon changing the model, we follow with a round of optimization and a fit over the full RoI. All parameters, such as the normalization and spectral parameters for all sources within $4^\circ$ of the GC, as well as for the diffuse components, are left free. We  search for additional point sources not included in the initial sky model and repeat the previous step.

Finally, we compute the SED of the GC {using the same energy binning as in the 4FGL-DR4 (but without any bin above 100 GeV)}. During this step, the normalizations of all background sources within  $4^\circ$ of the GC and diffuse model components are left free to vary.
{The final SEDs are reported in Appendix \ref{app:sed-on-off}.}

\section{Inference Methods}
\label{sec:inference}
In this study we work under the framework of Bayesian inference, with the aim of estimating the parameters of the luminosity function by leveraging SED data of GCs.\\
We describe first a classic likelihood-based technique, for which we explain how to calculate the likelihood and how to eventually use it for inference.\\
Next, we follow with a description of our SBI analysis, detailing the setup of our simulator and how the neural network is trained.

\subsection{Likelihood-based Bayesian analysis}
\label{sec:analysis-likelihood}


\subsubsection{Likelihood definition}
\label{sec:analysis-nested}
The likelihood is defined as $\log\mathcal{L}=-\frac{1}{2}\chi^2$, where the $\chi^2$ is computed as follows:
\begin{itemize}
    \item A set of parameters ($L_0$, $\sigma_L$, $N_1$,...,$N_{N_{GC}}$) is drawn from the prior as reported in Tab.~\ref{tab:priortable}. For each GC, { this will result in a real number for $N_1$,\ldots,$N_{N_{GC}}$. To effectively render these samples as integers, we round them to the next positive natural number $N_I$}
    ($I$ refers to the GC name { or an index assigned to enumerate them}, {$I=1,...,{N_{GC}}$ )}. { From now on, we use $N_I$ to refer to the individual number of MSPs in each GC. While the luminosity function (and thus its parameters $L_0$ and $\sigma$) is shared among all GCs, the numbers of MSPs for each GC $N_I$, are fully independent.}
    \item We draw individual MSP luminosities, ${L_c}$ ($c=1,...,N_I$), from the luminosity function and compute the GC SED in full analogy to the generation of synthetic data, cf.~Eqs.~\ref{eq:msp-norm}--\ref{eq:msp-cumul}. In particular, we work under the assumption that all MSPs follow a PLEC4 spectral profile with the average MSP parameters inferred from the 3PC, cf.~Sec.~\ref{sec:generate-synthetic}. 
     \item For each energy bin, if $F_{\rm MSP}^k>F_{\rm UL}^k$ (where $F_{\rm UL}^k$ represents the upper limit in that bin), the parameter set $(L_0,\sigma_L,N_I)$ is rejected.
    \item We define the following binned $\chi^2$ function: \begin{equation}
    \label{chi2}
    \chi^2=\sum_k\left(\frac{F_{\mathrm{MSP},I}-F_{\mathrm{GC},I}}{\Delta F_{\mathrm{GC},I}}\right)_k^2
    \end{equation}
    where $F_{\mathrm{GC},I}$ is the flux measured in the $I$-th GC with corresponding uncertainty $\Delta F_{\mathrm{GC},I}$ while $F_{\mathrm{MSP},I}=\sum_{c\in I}^{N_I}F_{\mathrm MSP,c}$ is the cumulative flux of all the (simulated) MSPs inside the $I$-th GC. The index $k$ runs over all energy bins {that are not upper limits}.
    \item For GCs whose emission is dominated by a single MSP, we include another contribution to the $\chi^2$, given as:
    \begin{equation}
    \chi_{\rm max}^2=\sum_k\left(\frac{(F_{\mathrm{MSP},I})_{\rm max}-F_{\mathrm{GC},I}}{\Delta F_{\mathrm{GC},I}}\right)_k^2
    \end{equation}
    where $(F_{\mathrm{MSP},I})_{\rm max}$ is the flux of the brightest MSP in the $I$-th GC. 
\end{itemize}
The prior is logarithmic for the reference luminosity and the number of MSPs for each GC, as detailed in Tab.~\ref{tab:priortable}. We assume a wide prior, between $10^{30}$ erg/s and $10^{40}$ erg/s for $L_0$ and between $0.01$ and $5$ for $\sigma_L$. For the number of MSPs inside every GC, we assume a range of $\sim1$ to $100$, except for GCs expected to host only a single MSP. In such cases, we reduce the maximum number to $10$, since the true value is unlikely to be higher.

\begin{table}[h]
    \centering
    \renewcommand{\arraystretch}{1.6}
    \begin{tabular}{c c c}
        \toprule
        Component & \qquad Parameter & Prior  \\
        \midrule
        \parbox[t]{1mm}{\multirow{2}{*}{\rotatebox[origin=c]{90}{lum fct.}}}
        & $L_0\;\left[\mathrm{erg}/\mathrm{s}\right]$ & $\operatorname{Log}\mathcal{U}(30, 40)$ \\
        & $\sigma_L$ & $\mathcal{U}(0.01, 5)$ \\
        \midrule
        \parbox[t]{1mm}{\multirow{2}{*}{\rotatebox[origin=c]{90}{GCs}}}
        & $N_I$ & $\operatorname{Log}\mathcal{U}(0, 2)$  \\
        & $N_{1, I}$ & $\operatorname{Log}\mathcal{U}(0, 1)$  \\
        \bottomrule
    \end{tabular}\textsf{'}
    \caption{Benchmark choices for prior ranges controlling the parameters of the MSP gamma-ray luminosity function and the number of MSPs per GC. If the measured gamma-ray emission of a GC has been demonstrated to arise from a single bright MSP, we lower the upper boundary of the respective prior, and indicate the number of MSP in these GCs as $N_{1,I}$. {Note that Terzan 5 receives an enlarged prior range $\operatorname{Log}\mathcal{U}(0, 2.7)$ (up to $\sim$500 MSPs) as the only exception from $N_I$ above.}
    Here, $\operatorname{Log}\mathcal{U}$ denotes the log-uniform distribution with base 10.
    }
    \label{tab:priortable}
\end{table}
To infer the parameters of the luminosity function in Eq.~\ref{eq:luminosity_function}, we maximize the likelihood using nested sampling, implemented through the \texttt{dynesty} python package (see \cite{2020MNRAS.493.3132S, sergey_koposov_2024_12537467}). We use $10^4$ live points (which leads \texttt{dynesty} to choose an evidence tolerance of $10.009$) and let \texttt{dynesty} select both the sample and bounding options. In particular, the algorithm selects uniform sampling with multiple bounding ellipsoids, given the number of free parameters of this study.

\subsection{Simulation-based inference analysis}
\label{sec:sbi}
In this work, we adopt SBI in its neural ratio estimation (NRE) flavor \citep{Miller:2022shs}. NRE recasts posterior inference as a supervised binary-classification problem: a neural discriminator $f_{\Phi}(x,z)$ {(with $x$ and $z$ being placeholder variables denoting generic data and parameters, respectively)} is trained to distinguish joint simulator draws $(x,z)\sim p(x,z)$ from independent draws $(x,z)\sim p(x)\,p(z)$. With equal class weights, the optimal classifier returns the sigmoid of the log density ratio, yielding an estimator of $r(z;x)\equiv \frac{p(z|x)}{p(z)}$, i.e.~the posterior-to-prior ratio (the ``likelihood-ratio trick'') \citep{Devroye:1996aa,Cranmer:2015bka}. The posterior then follows directly as $p(z|x)\propto r(z;x)\,p(z)$, without requiring an explicit likelihood. Since training data are generated by the forward simulator, nuisance parameters can be marginalized naturally and high-dimensional observables (here, gamma-ray SEDs) can be used without ad hoc data compression schemes beyond what is learned by the network.

We employ NRE as provided by the \texttt{swyft} package,\footnote{\url{https://swyft.readthedocs.io/en/stable/index.html}}, which jointly learns a task-specific summary statistic of the simulated data and the ratio estimator via the discriminator network $f_{\Phi}$ \citep{Miller:2021hys}. The computational gain arises because expensive likelihood evaluations are replaced by amortized network inference after training, making the overall approach scalable to large target sets \citep{Cranmer:2019eaq,2024arXiv240412484Z}.

In the remainder of this section, we outline our specific SBI workflow, which includes the simulator setup, network architecture and the associated training hyperparameters. We present the results of the SBI analysis on synthetic and real data in the following sections, while calibration and validation tests are described in App.~\ref{app:sbi-validation}.

\subsubsection{Simulator setup}
\label{sec:sbi-simulator}

The forward simulator of GC gamma-ray SEDs closely follows the approach outlined in Sec.~\ref{sec:analysis-likelihood}. A few parts need to be revised to achieve faster computations and compatibility with the SBI pipeline:
\begin{itemize}
    \item As explained in Sec.~\ref{sec:generate-synthetic}, synthetic GC SEDs are derived using {representative} MSP spectral parameters, thus we employ the same spectrum for all sources when examining those synthetic datasets. However, the simulation of GC SEDs changes for real LAT GC SEDs analysis with SBI {as in} this case, we extract the best-fit PLEC4 spectral parameters (see Eq.~\ref{eq:PLEC4}) from the list of 115 MSPs {contained} in 3PC. To each synthetic MSP, we randomly associate one of these spectral profiles and renormalize it such that the assigned luminosity $L_c$ is met. 
    \item For each energy bin, we redraw the assigned flux values $F_{k, I}$ by applying Gaussian noise:  $\tilde{F}_{k, I} \sim \mathcal{N}(F_{k,I}, \sigma_{k, \mathrm{4FGL}})$. Here $\sigma_{k, \mathrm{4FGL}}$ denotes the SED uncertainty reported in 4FGL-DR4 for the respective GC and energy bin $k$, which we average if upper and lower errors are asymmetric. {Should the redrawn flux $\tilde{F}_{k,I}$ be negative, we replace this value with $10^{-20}\,\mathrm{cm}^{-2}\,\mathrm{s}^{-1}$. In case 4FGL-DR4 states only an upper limit on $F_{k,I}$ in energy bin $k$, we assign $\tilde{F}_{k, I}$ a random number between the stated upper limit and $10^{-20}\,\mathrm{cm}^{-2}\,\mathrm{s}^{-1}$ drawn in log space.} 
    \item We account for the emission from multiple GCs by concatenating the spectra of each of them to obtain a one-dimensional data vector.
    \item Special treatment is given to those GCs that are dominated by the emission of a single MSP (NGC 6341, NGC 6652). While we allow for multiple MSPs to contribute to the total gamma-ray emission, we generate the total spectrum twice: The first time only using the brightest of the drawn MSPs and the second time adding all drawn MSPs (including the brightest one). Both spectra are added to the combined output data vector, counting this GC effectively twice. Whenever we include a GC with phase-resolved gamma-ray SEDs, we generate these two SEDs from the simulated population: The brightest of the drawn MSPs is associated with the ON-phase emission, whereas the cumulative emission of all remaining MSPs is taken as the respective OFF-phase SED. Both ON- and OFF-phase SEDs are appended to the final data vector.
\end{itemize}
By construction, the devised forward simulator utilizes measurement uncertainties derived from real LAT data. Hence, it is intrinsically trained on applying realistic noise for each GC based on its observed flux in each energy bin. Therefore, it is possible to infer model parameters from synthetic data and LAT GC SEDs alike. 

In addition, we highlight that the SBI pipeline is essentially trained on understanding the variety of GC gamma-ray SEDs (overall normalization and shape) as well as the bin-by-bin uncertainties arising from a dedicated gamma-ray analysis of each source. Using real data SEDs to define the simulator is not a source of genuine bias; it can rather be interpreted in a Bayesian way: We add prior knowledge about the zoology of real-sky GC SEDs and guide the method from the start.

\subsubsection{Training of the SBI method}
\label{sec:sbi-nn}

SBI requires training a discriminator network $f_{\Phi}$ to derive the ratio estimator from which the posterior distributions can be inferred. To do so, we exploit the functionalities of the adopted software \texttt{swyft} and the neural network definitions it provides. All details about the network architecture are given in App.~\ref{sec:sbi-nn}, as well as the definition of our training hyperparameters in terms of training data sizes, batch sizes, number of epochs, etc. The training is conducted on a single A100 GPU with 40 GB RAM.

\section{Simulations: Results}
\label{sec:sims-results}

\subsection{Likelihood-based Bayesian analysis}
\label{sec:analysis-nested-results}

\begin{figure*}
    \centering
    \includegraphics[width=0.49\linewidth]{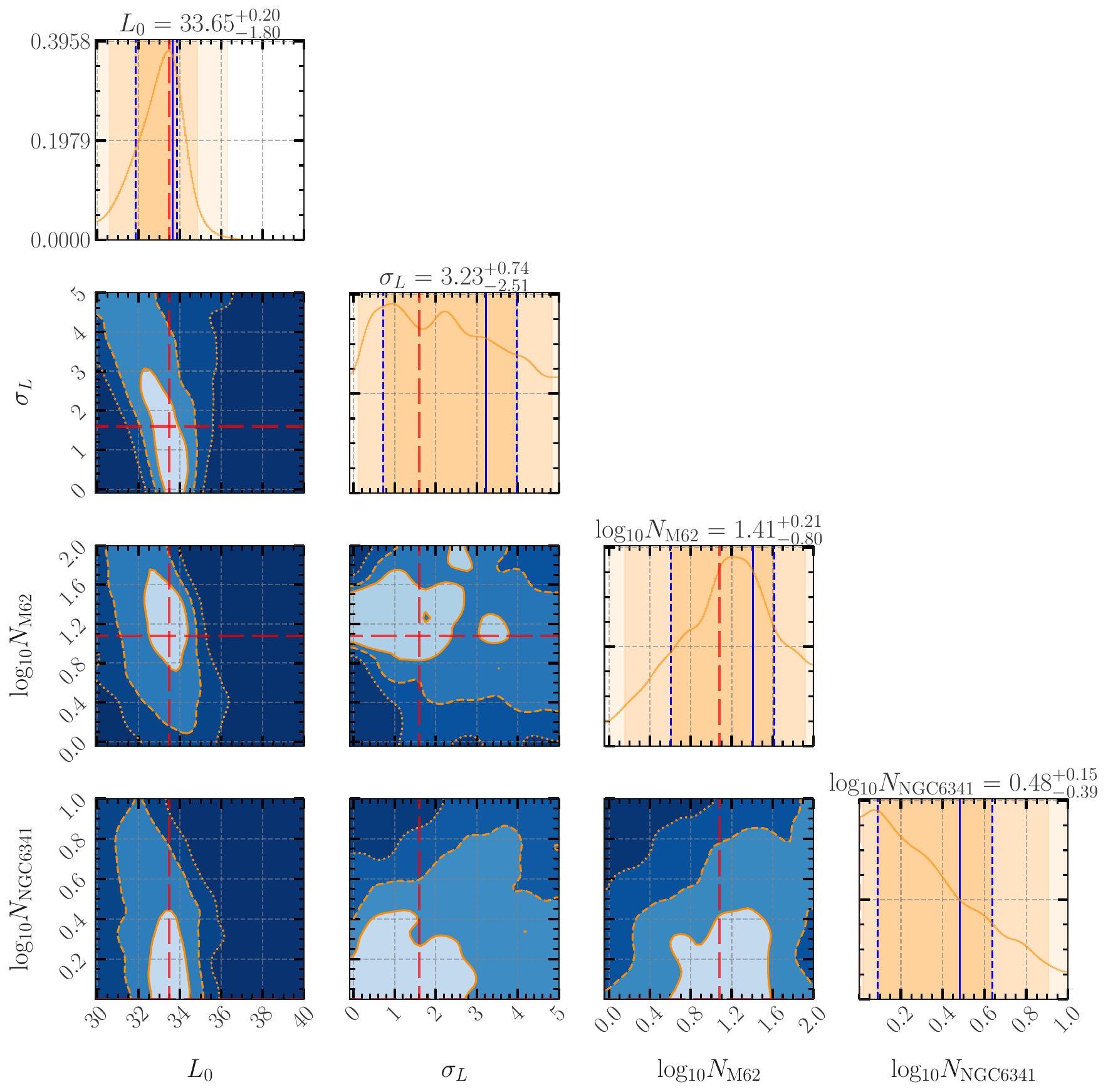}\hfill
    \includegraphics[width=0.49\linewidth]{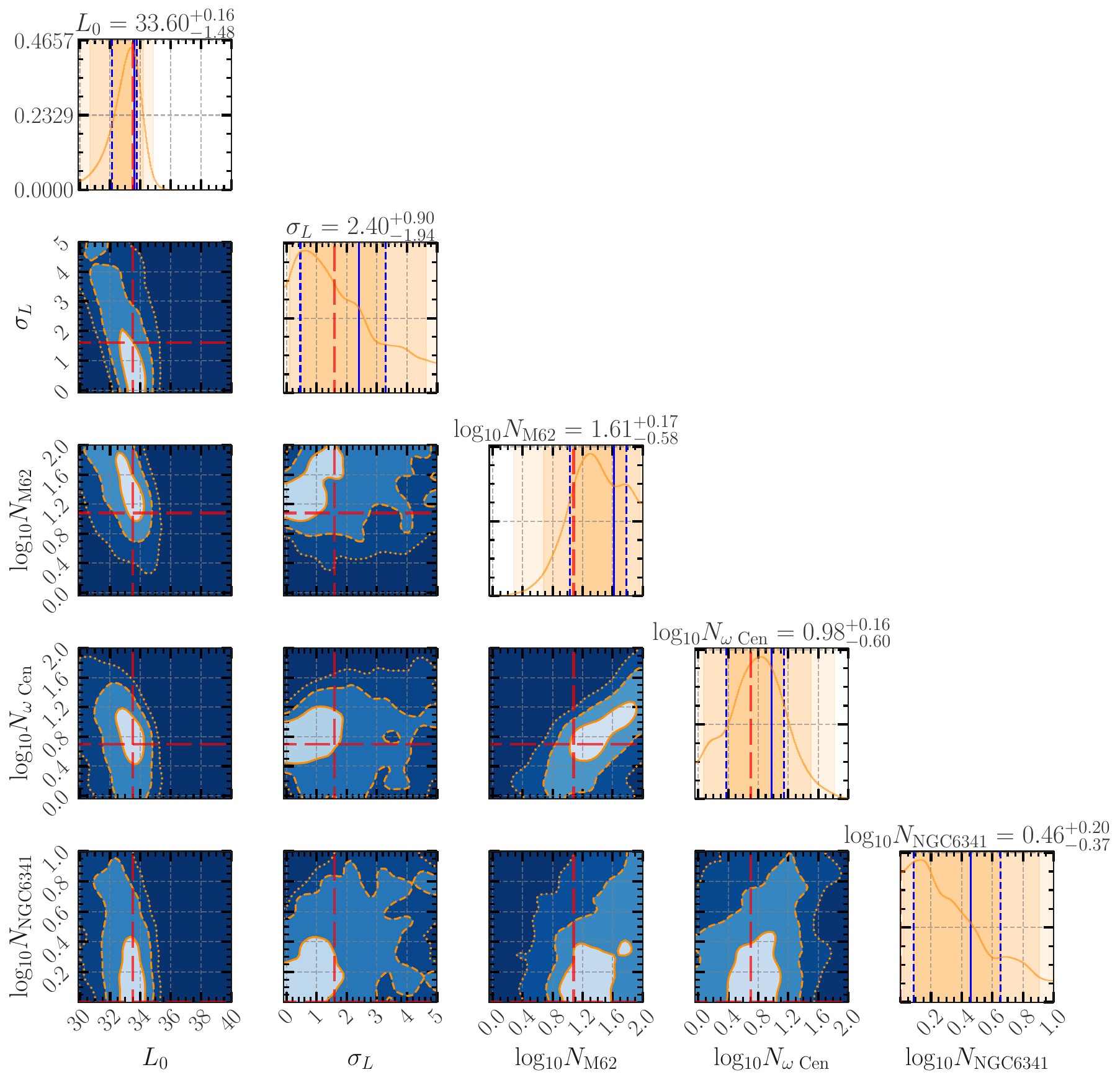}
    \caption{Marginal posteriors for synthetic GC SEDs from the likelihood-based analysis. The dashed red lines represent the injected true values, while the solid blue lines represent the median recovered values. For clusters with only 1 {MSP}, the true value {of $N_I$} corresponds to the lower extreme {($\log{N_I}=0$)}. The vertical dashed blue lines in the diagonal plots delimit the $1\sigma$ upper and lower uncertainties (both uncertainties and the median are also indicated in the panel title). The orange shaded regions denote the $3\sigma/2\sigma/1\sigma$ credible bands, listed in order of increasing opacity. \textbf{Left:} Case of 2 GC SEDs: M 62 and NGC6341. \textbf{Right:} Case of 3 GCs SEDs: M 62, $\omega$-Centauri and NGC6341.}
    \label{fig:results-dynesty}
\end{figure*}

First, we explore the performance of the likelihood-based method using the synthetic GC SED dataset (see Sec.~\ref{sec:generate-synthetic}). 
In Figure \ref{fig:results-dynesty}, we show two corner plots, detailing our results applied first to 2 regular GCs, with no information on their numbers of MSPs (M 62 and NGC 6341, left panel) and then 3 GCs, adding 1 GC where the emission coincides with a known MSP shining towards Earth from the GC direction ($\omega$-Centauri, right panel). The diagonal of the corner plots reports the one-dimensional posterior distributions (solid orange line) of each model parameter marginalized over the remaining parameters. The solid blue line indicates the median value of the parameter, whereas the dashed blue lines mark the boundaries of the 68\% credible interval around the median. These values are also explicitly quoted in the title of each panel. Additionally, the shaded orange bands denote the 3$\sigma$/2$\sigma$/1$\sigma$ credible bands distinguished by increasing opacity. In contrast, the sub-diagonal panels show the two-dimensional marginal posterior distributions for parameter tuples providing a measure of mutual correlation. 
With red dashed lines, we indicate the {true} values for all parameters, while the blue-shaded regions denote the $3\sigma/2\sigma/1\sigma$ credible bands, listed in order of increasing opacity. Looking at the left panel, we see that for both $L_0$ and $\sigma_L$ their true values lie in the $1\sigma$ band (also delimited by the blue dotted vertical lines in the diagonal plots). In particular, we observe that $L_0$ is well recovered, with the true value being $3\times10^{33}$ erg/s and the recovered median one (corresponding to the blue solid line) being $\sim4\times10^{33}$ erg/s. 
We also notice that we are able to exclude values above $\sim3\times10^{36}$ erg/s. This does not necessarily apply to the numbers of MSPs in each GC, as can be seen for NGC 6341. For this GC, only 1 MSP was injected (as seen in Section \ref{sec:generate-synthetic} and in Table \ref{tab:gc-sets}) and it was treated as explained in section \ref{sec:analysis-likelihood}. However, its injected value is only recovered at $2\sigma$ for both the left and right panels. 
Adding 1 GC decreases the width of the $1\sigma$ credible band for all parameters ($L_0$ sees a decrease of $\sim18\%$ and $\sigma_L$ sees one of $\sim14\%$) while also shifting the median recovered values closer to the injected ones (this is especially true for $\sigma_L$, whose median value goes from 3.23 to 2.40 while its true value is of 1.6 as indicated in Table \ref{tab:gc-sets}). Looking at $L_0$, we see that we are now able to exclude values above $\sim10^{35}$ erg/s, while for the number of MSPs in M62 we see that we can now exclude values below $\sim2$ MSPs. For the newly added $\omega$-Centauri, our method allows for the exclusion of numbers above $\sim63$ MSPs.

As explained so far, the inclusion of 2 GCs allows to reproduce the injected parameters, even if with associated large posteriors. Increasing the number of GCs seems to improve our results, however we find ourselves unable to further add more GCs, falling into the curse of dimensionality.
Looking at the CPU time  required to reach an evidence of $d\log\mathcal{Z}=10$ for the same computational setup, corresponding to 32 processes run in parallel on a common computer (Lenovo Thinkpad L14 G4, mounting a 6 cores AMD Ryzen 5 PRO 7530U and 32 GBs of RAM), we find that $\sim26$ minutes are required for 1 GC (M 62). At the same time, for 2 GCs (M 62 and $\omega$-Centauri) this time increases to $\sim276$ minutes. We repeat this analysis adding one GC with 1 MSP (NGC 6341) and find an even greater increase in CPU time (this time approximately $\sim952$ minutes are required to reach an evidence of $d\log\mathcal{Z}=20$).
This increase in complexity and computational time prevented us from using this method with more than 3 GCs, which means that we could not fully exploit the synthetic GC sample, nor use this approach to analyze real data.

We conclude that the traditional likelihood-based nested sampling approach we implemented starts to quickly become intractable and the sharp increase in computation time prevents us from reaching our objectives and any further exploration of the problem. Nested sampling is thus not found to be the most adequate method of choice to address our physics question.  A suitable remedy to lower the computational costs of the analysis is the use of more efficient methods like the SBI technique introduced in Sec.~\ref{sec:sbi}. 

\subsection{Comparing simulation-based inference with the likelihood-based method}
\label{sec:sbi-results-dynesty}

\begin{figure}
    \centering
    \includegraphics[width=\columnwidth]{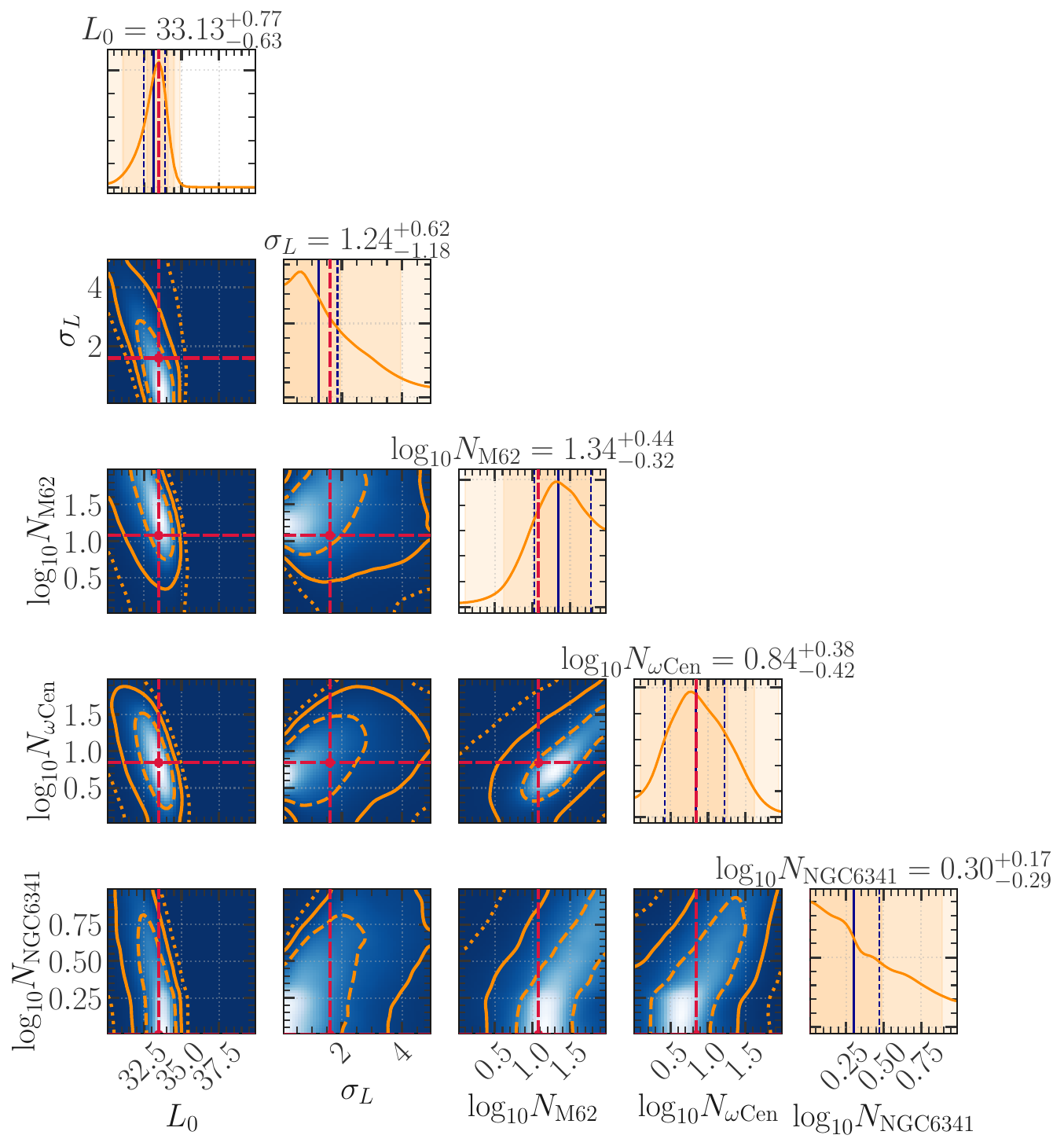}
    \caption{Same as the right panel of Fig.~\ref{fig:results-dynesty} using the SBI approach to derive the two-dimensional marginal posteriors for synthetic GC SEDs. 
    }
    \label{fig:SBI-vs-dynesty}
\end{figure}

In light of the computational hurdles that the likelihood-based nested sampling approach faces, we first consider applications to synthetic GC target datasets and revisit the GC target lists we already explored in the previous section. We visualise our findings in Fig.~\ref{fig:SBI-vs-dynesty} focusing on the GC selection in the right panel of Fig.~\ref{fig:results-dynesty}.

Since the underlying prior definitions are identical, we can directly compare likelihood-based and SBI posteriors per parameter. In fact, the likelihood-based posteriors are a litmus test for the SBI approach since posteriors derived via implicit methods are asymptotically consistent. This means that such posteriors converge to the true Bayesian posterior $p(z|x) \propto p(x|z)p(z)$ that would be obtained from the exact likelihood, whenever the latter is defined.

The SBI posteriors are consistent with their likelihood-based counterparts, as they exhibit similar shapes and behaviours near the prior edges. We interpret this as confirmation that Bayesian SBI can be safely applied to this inference problem even beyond the number of GCs accessible via the likelihood-based method. 
Additionally, we verified the well-calibration of the SBI posteriors using Bayesian coverage tests, which showed that they are neither overconfident nor overly conservative. The details are provided in App.~\ref{app:sbi-coverage}.

Regarding the computation time for each set of posteriors, we found that the case with two GCs takes about 8 hours and 40 minutes, while the case with three GCs finished after approximately 11 hours on the GPU mentioned above. First and foremost, the SBI approach does not generate a massive increase in computation time when enlarging the set of considered GCs. Secondly, these numbers need to be taken with a grain of salt since they depend on many of our training hyperparameter definitions for the baseline run. The number of one million training samples is very conservative. We checked that using only $10^5$ samples reduces the computation time for the 2-GC case to about 51 minutes (3077 s) while essentially producing the same well-calibrated posteriors. Additionally, our setup of learning rate, learning rate scheduler, and early stopping patience is likewise conservative. Optimizing these hyperparameters for time could potentially speed up the analysis even more. However, in comparison to the week-long computation time of the likelihood-based analysis {when considering 4 GCs at the same time}, the SBI approach is already much better in this baseline run. 

\subsection{Combining more globular clusters}
\label{results:sbi-synthetic-more-sources}

Since, with the SBI framework, we are not constrained by computing time any longer, we now analyze in more depth the full sample of synthetic GCs. We stress that all of these synthetic SEDs represent phase-averaged SEDs. We start by considering those six GCs for which we do not impose any constraint regarding the number of already detected gamma-ray MSPs (the first six in Table \ref{tab:synthetic-SEDs}).  While this dataset increases the number of GCs, it also erases any constraint on the expected number of MSPs per GC that NGC 6341 did provide in the previous scenarios discussed in Secs.~\ref{sec:analysis-nested-results} and~\ref{sec:sbi-results-dynesty}. Hence, we can quantify the impact of external information that act as a prior on $N_{I}$. To ease the comparison with the early case of three GCs, we show the two-dimensional marginal posterior distributions for only those parameters that are shared between the GC sets in the left panel of Fig.~\ref{fig:SBI-numberConstraints}. The full inference results are reported in the appendix in Fig.~\ref{fig:SBI-6GC-noConstraints}.

While the pivot luminosity $L_0$ can be largely recovered from the data as shown in Fig.~\ref{fig:SBI-numberConstraints}, the width of the log-parabola stays widely unconstrained at approximately the same level that only the three GCs in Fig.~\ref{fig:SBI-vs-dynesty} achieved. Moreover, $\sigma_L$ runs against the lower boundary of its prior range, i.e., it is biased towards narrow luminosity functions. In addition, all recovered $N_I$ are biased towards the upper boundary of the prior range, sometimes without including the true value within the 95\% credible band. This is a natural consequence of $\sigma_L$ being too narrow given the true values. This is in stark contrast to Fig.~\ref{fig:results-dynesty} where NGC 6341 adds a prior on the number of MSPs and the $N_I$ are recovered within the 68\% credible interval. We conclude, first, that adding these constraints is essential to render the entire approach faithful. Secondly, simply adding more GCs to the dataset without any guidance on the number of MSPs in at least one GC does not improve the inference quality per se.

\begin{figure*}
    \centering
    \includegraphics[width=0.49\linewidth]{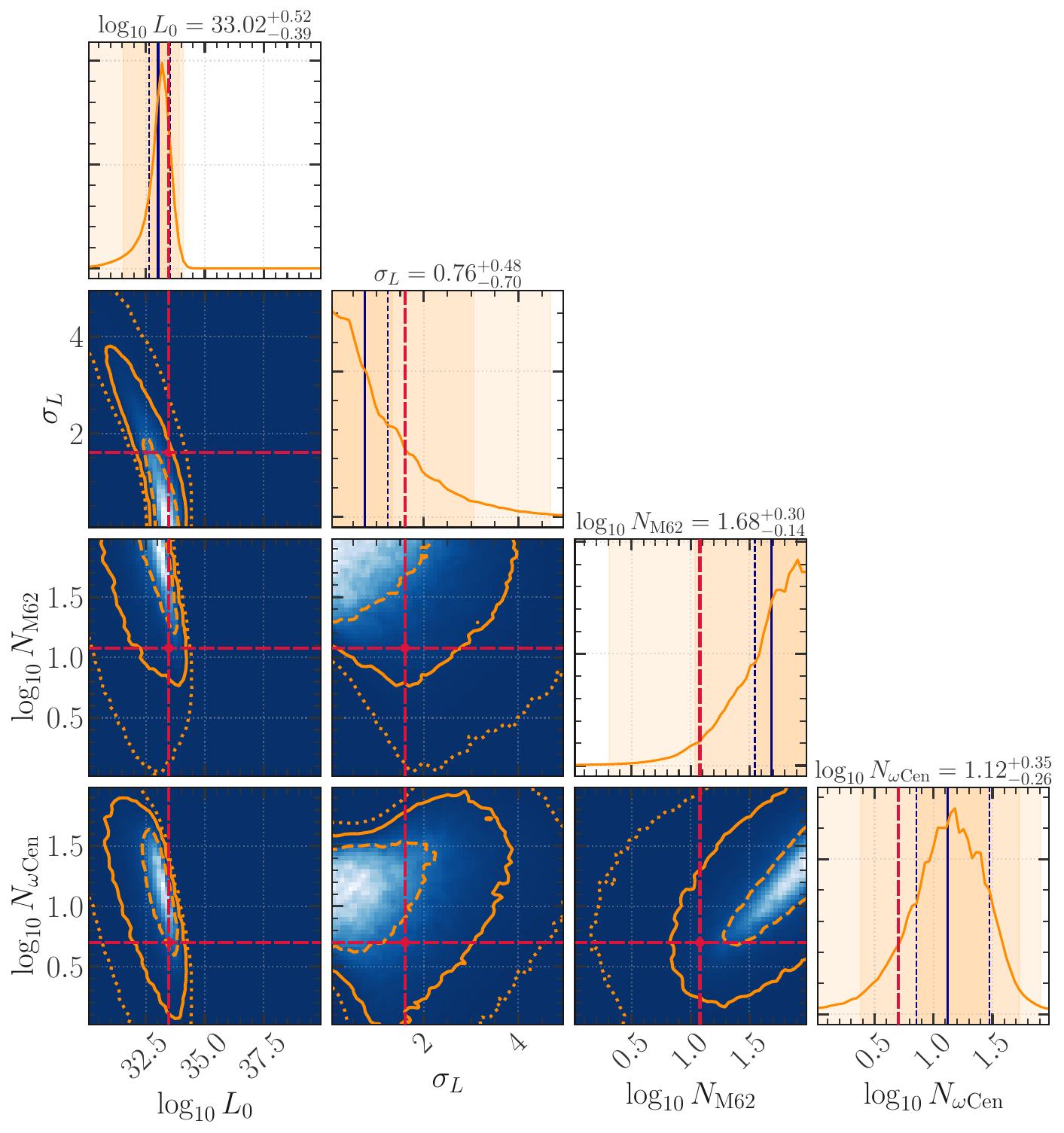}\hfill
    \includegraphics[width=\columnwidth]{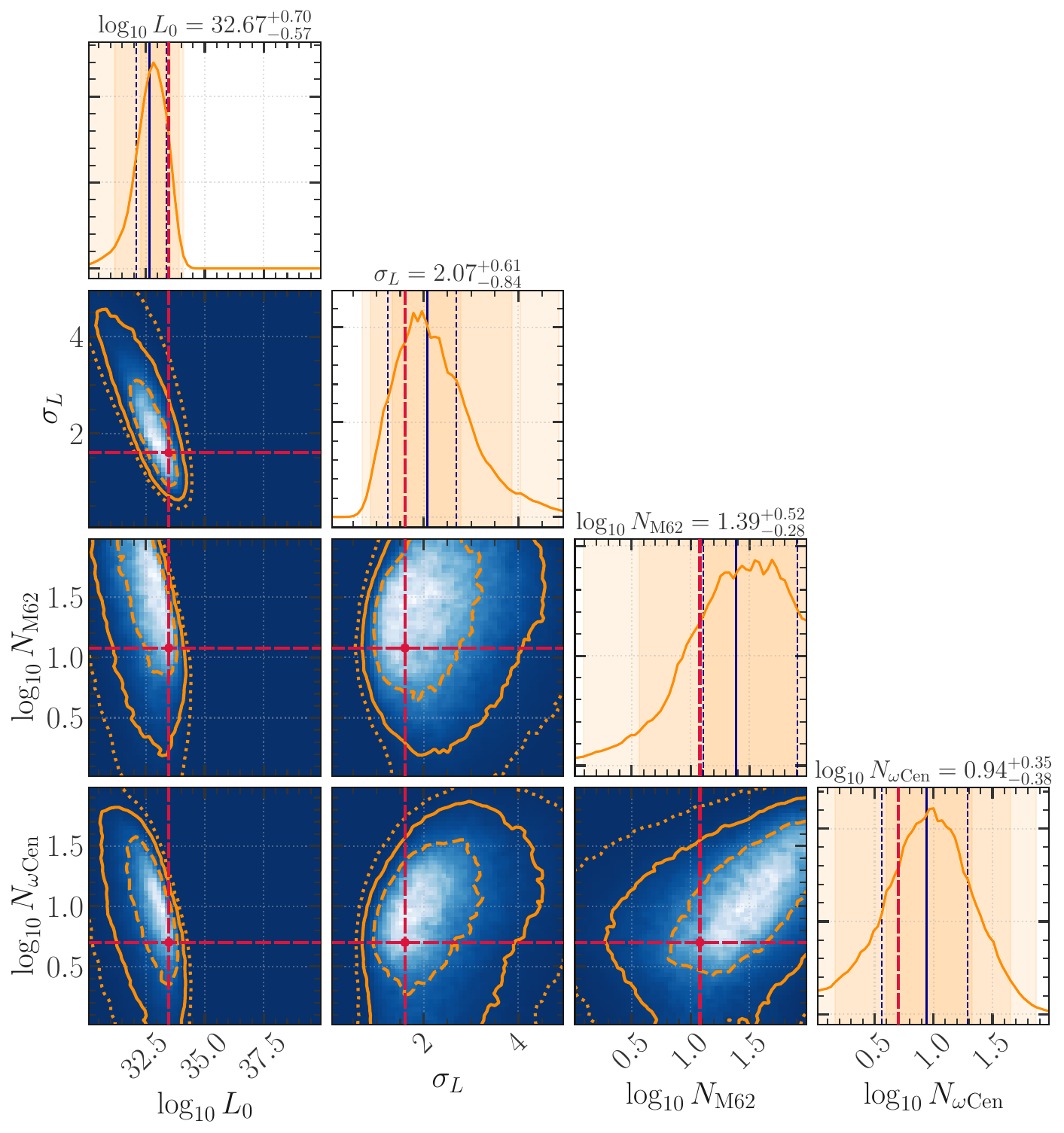}
    \caption{Selected two-dimensional marginal posteriors for different synthetic GC datasets derived with the SBI approach. \textbf{Left}: First six GCs in Tab.~\ref{tab:synthetic-SEDs}; providing no constraint on the expected number of MSPs in any of the clusters; \textbf{Right}: Adding NGC 6341 and NGC 6652 to these six GCs and thereby adding such MSP number constraints.
    }
    \label{fig:SBI-numberConstraints}
\end{figure*}

As a next step, we add NGC 6341 and NGC 6652 to the set of six GCs and, thus, gain the constraint that their observed emission is due to exactly one MSP. The obtained two-dimensional marginal posterior distributions for the same selected parameters are shown in the right panel of Fig.~\ref{fig:SBI-numberConstraints}, whereas the full triangle plots featuring all parameters are in the appendix in Fig.~\ref{fig:SBI-allGCs-noONOFF}. First, we observe that the overall quality of the inference is satisfactory since six of the ten inferred parameters are recovered within the 68\% credible interval of their respective posterior. Second, we display the coverage test results for this setup in Fig.~\ref{fig:coverage-8GC} of App.~\ref{app:sbi-coverage}. The coverage of all posteriors is very close to optimal calibration, implying that our inference is faithful.

Fig.~\ref{fig:SBI-numberConstraints} underlines our earlier statement that solely increasing the number of GCs in the dataset does not necessarily improve the sharpness of the posteriors; note the difference between the posteriors of $\sigma_L$, in particular. What brought us most of the improvement was adding the right GCs, namely those that bear a constraint on their total MSP content. In fact, with such constraints, we are able to robustly exclude values $\sigma_L < 0.5$ at more than $3\,\sigma$ while we could only set upper limits before.

However, the number of considered GCs does play a role in the analysis. The set of three GCs including NGC 6341 in Fig.~\ref{fig:SBI-vs-dynesty} allows us to only set upper limits on $\sigma_L$. That changes when the GC sample size grows. We therefore conclude that adding more GCs to the target GC set is highly beneficial and warranted to constrain the gamma-ray luminosity function of MSPs in GCs. In contrast, the individual numbers of MSPs per GC $N_I$ are not strongly constrained, even with eight GCs in the target dataset. This observation is not surprising since the two-dimensional marginal posteriors demonstrate that the individual GC parameters are not strongly correlated with each other but are more significantly correlated with the parameters of the luminosity function. Therefore, the ensemble of GCs (and the uncertainty of their individual MSP populations) is the source of the improvement we see in the inference of $L_0$ and $\sigma_L$. It is not a single GC that drives the precision of the inferred parameters.

\subsection{The role of phase-resolved gamma-ray spectra}
\label{results:sbi-synthetic-ONOFF}
At this point, our exploration of GC gamma-ray SEDs to infer the GLF of MSPs has not leveraged a last scenario that may occur in those GCs with a confirmed gamma-ray-bright MSP. As outlined in Sec.~\ref{sec:gc-sed}, it is possible to exploit the phase information of the detected MSP to separate photon events that coincide with the ON-phase (MSP's beamed emission directed towards us) and the OFF-phase of the MSP. While the ON-phase emission is predominantly related to the single MSP, the OFF-phase emission -- if significant -- must be the result of the cumulative gamma-ray flux of all remaining MSPs of the respective GC's population. Therefore, combining ON- and OFF-phase SEDs from the same GC equips us with very sensitive observables to constrain the number of MSPs in the target. Here, we now examine the impact of including such phase-resolved SEDs in our SBI approach to infer the underlying model parameters. 

To this end, we utilize the synthetic ON- and OFF-phase data of NGC 6624 that we prepared in Sec.~\ref{sec:generate-synthetic}. This cluster is added to the target set of GCs analyzed in the previous section. The obtained two-dimensional marginal posterior distributions are displaced in Fig.~\ref{fig:SBI-allGCs-wONOFF}, where the last row of the triangle plot refers to the results related to NGC 6624. The quality of the inference remains good with eight injected parameters being reconstructed within the 68\% credible interval of their respective posterior. Likewise, the coverage test in Fig.~\ref{fig:coverage-9GC} of App.~\ref{app:sbi-coverage} confirms the posteriors' calibration close to optimum. The effect of the phase-resolved SEDs is quite evident, NGC 6624 exhibits constraining power on $L_0$ and $\sigma_L$ at the level (or even higher) than M 62, the GC with the largest inferred MSP population. As a consequence, the posterior of $\sigma_L$ is narrowed down even further, and the number of sources in NGC 6624 is among the best-defined posteriors. We therefore conjecture that stacking many GCs with phase-resolved gamma-ray emission would constitute the ideal approach to explore the MSP GLF from SEDs. Unfortunately, only two cases have been reported in the literature so far. 


\section{Fermi-LAT data analysis: Results}
\label{results:sbi-LATdata}
Having established and demonstrated the general constraining power of the GC SED analysis regarding the MSP luminosity function's parameters via synthetic data, we now apply the approach to real gamma-ray SEDs of detected GCs. 

As a reminder, when transitioning to real LAT SEDs, we also make our SED simulations more complex by allowing for variable MSP spectra informed by observed MSPs, rather than assuming that each MSP follows {the representative} spectrum. 

We will adopt the GC target datasets described in Sec.~\ref{sec:gc-samples} with SED as obtained in Sec.~\ref{sec:gc-sed}. We therefore leverage all GCs hosting individually detected MSPs, using this fact as a constraint or using the phase-resolved (ON \& OFF) SEDs.

\subsection{Luminosity function results with baseline target data}
\begin{figure*}[t]
    \centering
    \includegraphics[width=\columnwidth]{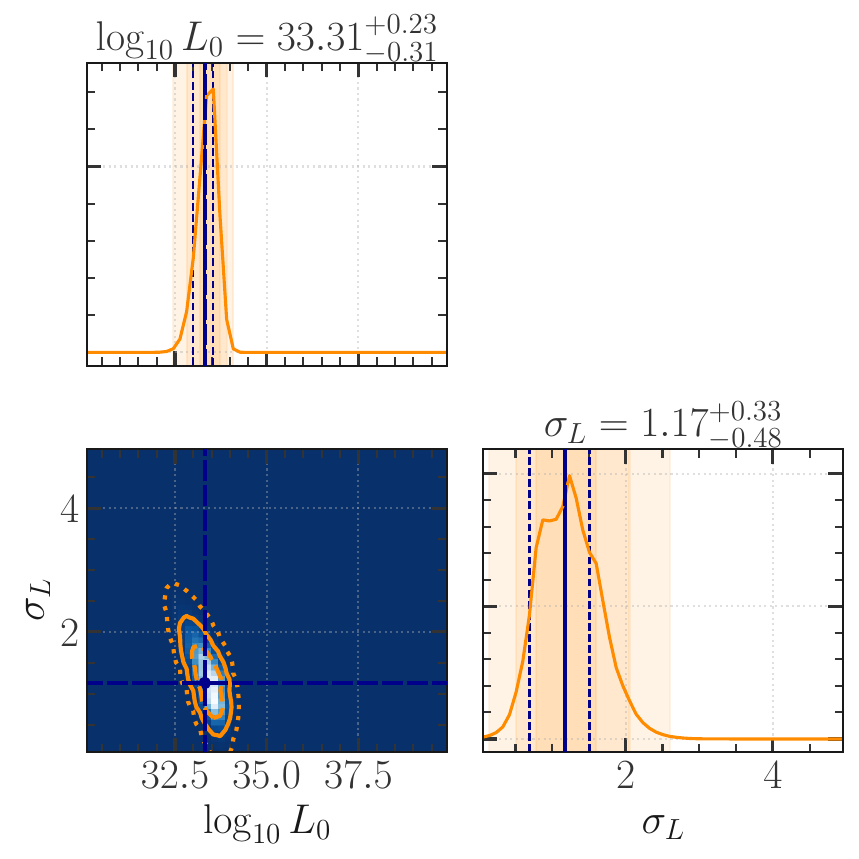}\hfill
    \includegraphics[width=0.9\columnwidth]{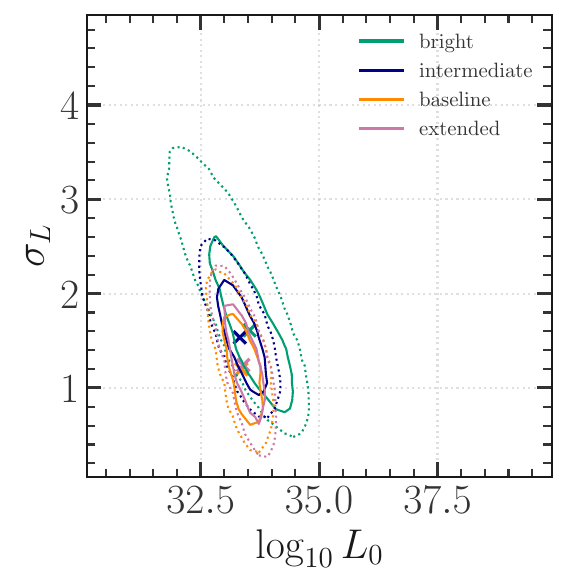}
    \caption{\textbf{Left:} Two-dimensional marginal posterior distributions obtained using the SBI approach on real Fermi-LAT data solely targeting the MSP GLF parameters $L_0$ and $\sigma_L$ leveraging the baseline target data. The thick dashed blue lines indicate the mean inferred parameter values, while the $1\sigma$ upper and lower uncertainties are represented by vertical dashed blue lines in the respective diagonal panels. The {dashed/solid/dotted} orange lines refer to the 1$\sigma$/2$\sigma$/3$\sigma$ credible intervals of the marginal two-dimensional posterior. \textbf{Right:} The same Two-dimensional marginal posterior distribution inferred for target samples of increasing size: bright (11; green), intermediate (25; blue), baseline (36; orange), and extended (44; magenta) target data. Solid (dotted) contours denote the 68\% (95\%) credible regions, and crosses indicate the posterior median estimates.}
    \label{fig:SBI-real-L0sL}
\end{figure*}

It is a feature of the NRE flavor of the SBI method to render accessible marginal posteriors of interest without the need to infer the full joint posterior. Therefore, we can marginalize over all parameters $N_I$ that yield the number of MSPs per GC (see also the discussion below) and solely infer the one- and two-dimensional marginal posterior distribution for $(L_0, \sigma_L)$. By employing the identical training and inference strategy for our SBI method as for the case of synthetic data, we derived the posteriors for $L_0$ and $\sigma_L$ from the baseline target data. Our findings are shown in the left panel of Fig.~\ref{fig:SBI-real-L0sL}. 

In this setup, we find informative posterior distributions that exclude a sizeable portion of the assumed prior range. We obtain $\log_{10}L_0=33.4^{+0.4}_{-0.2}$ and $\sigma_L=1.3^{+0.4}_{-0.5}$ as the median predictions (and 68\% credible bands) for the MSP luminosity function's parameters. The achieved precision of these posteriors is better than the one we were able to derive in the case of synthetic data and nine GC SEDs featuring only one GC with ON- and OFF-phase measurements (see Fig.~\ref{fig:SBI-allGCs-wONOFF}). The reason for this improvement is very likely the increase in the total number of considered GCs by a factor of four compared to the test case with synthetic data. 

\subsection{Impact of the size of the target dataset} 
The right panel of Fig.~\ref{fig:SBI-real-L0sL} overlays the two-dimensional marginal posterior distribution of $L_0$ and $\sigma_L$ obtained from the four different target data sets growing in size of the considered GCs. The marginal posteriors exhibit a clear trend towards convergence and tightening as the target sample size is enlarged from the bright (11 clusters) to the intermediate (25), baseline (36), and extended (44) target datasets. For the bright-only target set, the constraints are comparatively weak and strongly elongated in the $\sigma_L-\log_{10}L_0$, revealing a pronounced anticorrelation: models with a lower pivot luminosity $L_0$ can be compensated by a broader luminosity function (larger 
$\sigma_L$) to reproduce the high-luminosity tail that dominates a sample of mostly gamma-ray bright GCs. As progressively fainter clusters are included, this degeneracy is increasingly broken and the credible region contracts, with the posterior mass shifting mildly towards higher $L_0$ and smaller $\sigma_L$. This is consistent with the requirement to simultaneously fit both the bright and faint ends of the GC populations. { The preceding statement manifests itself as follows: The baseline selection of GCs adds mainly dim clusters to the intermediate target set. If we were to explain the baseline set with the inferred LF parameters for the intermediate sample, we would predict, on average, more MSPs in the bright and faint tails of the LF. Such tails cannot exist to explain the diversity of the gamma-ray emission observed in these added dim GCs, i.e., (on average) one would expect too often a too faint or too bright GC. Therefore, the LF must be narrower, whereas the average luminosity $L_0$ can almost stay the same. In this sense, it is the mentioned diversity of the level of gamma-ray emission observed in the Milky Way's GCs that breaks the degeneracy and informs the gamma-ray LF parameters in this method.}

Importantly, the baseline and extended posteriors overlap closely, indicating that the inference has largely converged by using about 36 clusters. Adding the remaining targets (most of which are rather faint and have not been firmly associated with a GC in 4FGL-DR4) primarily improves precision rather than causing a noticeable shift in the preferred luminosity-function parameters. Hence, we conclude that it is fully sufficient to consider the baseline target data set to obtain stable results.

\subsection{Number of gamma-ray MSPs in selected globular clusters} 
\begin{figure}[t]
    \centering
    \includegraphics[width=\columnwidth]{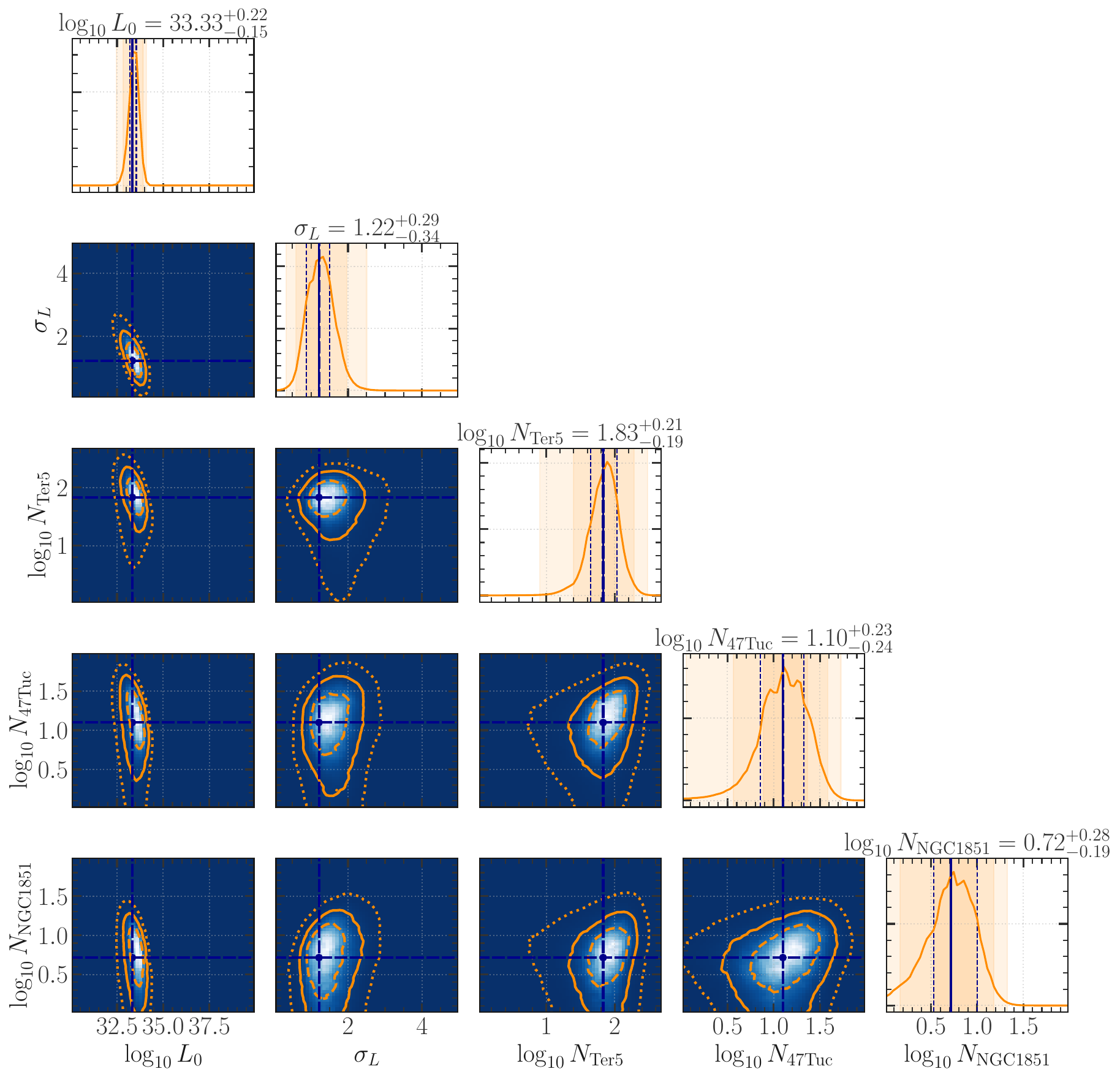}
    \caption{Two-dimensional marginal posteriors for the pivot luminosity $L_0$ and log-normal width $\sigma_L$ of the MSP gamma-ray luminosity function, jointly inferred with the cluster-specific (integer) MSP counts for Terzan 5, 47 Tucanae, and NGC 1851 using the baseline target dataset (36 globular clusters). Solid (dashed, dotted) contours indicate the $1\sigma$ ($2\sigma$, $3\sigma$) credible regions; crosses indicate the medians of the posteriors.}
    \label{fig:SBI-real-NMSP}
\end{figure}

As stated above, we can use the NRE flavor of the SBI method to select parameters of interest. Now, we want to examine whether we can reasonably constrain the number of gamma-ray-bright MSPs in a small set of GCs. To this end, we choose Terzan 5, 47 Tucanae and NGC 1851, motivated by the fact that these GCs feature substantial populations of MSPs detected in the radio waveband\footnote{See also the overview website curated and actively updated by P.~Freire: \href{https://www3.mpifr-bonn.mpg.de/staff/pfreire/GCpsr.html}{Pulsars in globular clusters}.} \cite{Ridolfi:2021idl, Ridolfi:2022gmc, Padmanabh:2024bsz}. Using the baseline target dataset, we obtain joint constraints on the luminosity function's parameters and the cluster-specific MSP counts displaced in Fig.~\ref{fig:SBI-real-NMSP}. The marginal posteriors yield for the  MSP counts $\log_{10}N_{\rm Ter5} = 1.8^{+0.2}_{-0.2}$, $\log_{10}N_{\rm 47\,Tuc} = 1.1^{+0.3}_{-0.2}$, and $\log_{10}N_{\rm NGC\,1851} = 0.7^{+0.3}_{-0.3}$. The posterior central values imply an ordering $N_{\rm Ter5} > N_{\rm 47\,Tuc} > N_{\rm NGC\,1851}$, with only partial overlap of the corresponding credible intervals, indicating that the data prefer different MSP normalizations for these three clusters within the adopted model. The posterior mass is well confined within the prior bounds for all parameters, and the marginals are unimodal, indicating that the data contains enough information under the assumed model and that the parameters are identifiable at the level shown by the credible regions. In Appendix \ref{app:prior-impact}, we examine the impact of changing the assumed prior profile of all parameters $N_I$. In particular, we move from a log-uniform distribution to a linear-uniform distribution over the same range of values and observe that the resulting posteriors predict a larger MSP count per GC by a factor of a few and a reduced $L_0$ but larger $\sigma_L$. While this consistency check reveals a non-negligible prior dependence, we note that a log-uniform {distribution is ``scale-free'' and thus suitable in cases where a parameter's order of magnitude is largely unknown. Hence, our results under this assumption} are rather agnostic about the expected MSP counts per GC and are a more natural choice for this sort of problem.

The joint posterior in Fig.~\ref{fig:SBI-real-NMSP} also exhibits the expected covariance structure between cluster-level and population-level parameters. In particular, each $N_{I}$ is anti-correlated with $\log_{10}L_0$ and $\sigma_L$, reflecting the trade-off between the number of sources and the typical (or dispersed) per-source luminosity required to reproduce the cumulative GC gamma-ray emission. The cluster-specific $N_I$ parameters are mutually positively correlated, as they are coupled through the shared luminosity function parameters.

\section{Discussion}
\label{sec:discussion}


\subsection{Comparison with previous determinations of the MSP gamma-ray luminosity function}
\label{sec:iterature-comparison}

\begin{figure}[t]
    \centering
    \includegraphics[width=\columnwidth]{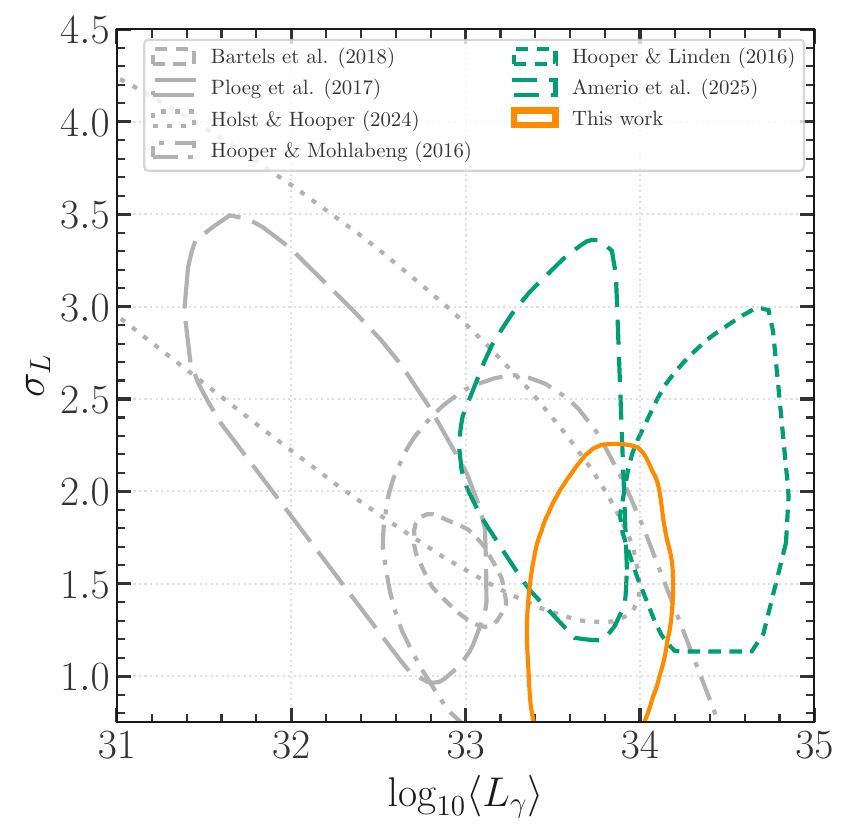}
    \caption{Comparison of 95\% credible/confidence regions in the $(\log_{10}\langle L_\gamma\rangle,\sigma_L)$ plane for a log-normal MSP gamma-ray luminosity function, where the mean luminosity is defined as $\langle L_\gamma\rangle = L_0 \exp(\sigma_L^2/2)$. The solid orange contour shows the joint 95\% credible region obtained in this work (baseline target dataset of 36 globular clusters), while dashed contours indicate literature constraints. The colored contours refer to works targeting MSPs in GCs (\cite{Hooper:2016rap}, \cite{Amerio:2024qor}), whereas grey contours refer to works addressing field MSPs (\cite{Hooper:2015jlu}, \cite{Ploeg:2017vai},  \cite{Bartels:2018xom},  \cite{Holst:2024fvb}).}

    \label{fig:literature-comparison}
\end{figure}

Figure~\ref{fig:literature-comparison} places our MSP GLF posterior inferred from the baseline target dataset in the context of representative results from the literature. For ease of comparison across different parameterizations, we express our inference in terms of the mean luminosity of a log-normal luminosity function,
\begin{equation}
    \langle L_\gamma \rangle \;=\; L_0\,\exp\!\left(\frac{\sigma_L^2}{2}\right),
\end{equation}
and report (as in the figure) 95\% credible intervals for our posterior and 95\% confidence/credible regions for the reference studies, as quoted by the respective works. 
We emphasize that a direct, one-to-one comparison of results from GCs analyses and ``field'' MSP ones is only meaningful under the hypothesis that intrinsic luminosity properties are shared among these environments.
The contours in grey refer to results on field/disc MSPs. We stress that any comparison of works is subject to systematic effects because of differing GC/MSP sample definitions, distinct data products (integrated fluxes versus spectral information), and distinct treatments of distances, selection effects, and additional correlations. Consequently, differences between contours should not be interpreted solely as astrophysical discrepancies; they may also reflect modeling and data-selection choices.

\paragraph{Consistency and interpretation of Fig.~\ref{fig:literature-comparison}.}
With the above differences in mind, Fig.~\ref{fig:literature-comparison} shows that our 95\% credible region occupies a parameter space domain that overlaps with the broad envelope spanned by existing findings in the literature, and in particular is consistent with the range of GC-based constraints within the quoted uncertainties. At the same time, the dispersion among field-MSP contours is itself non-negligible, underscoring that catalog choice (e.g.~3FGL versus 3PC{, corresponding respectively to 4 years and 12 years of {\it Fermi-LAT} data}), the treatment of distance information, and the modeling of selection effects can shift or elongate the inferred region in the $(\log_{10}\langle L_\gamma\rangle,\sigma_L)$ plane. Therefore, the primary message of this comparison is not that all studies must agree at the level of their best-fit points, but that there exists a broad region of mutual support once one accounts for differences in target population and methodology.

These considerations extend to the potential consequences of the inferred GC-based luminosity function for the properties 
of the bulge MSP population hypothesized to account for the Galactic Center GeV Excess. Our inferred GC luminosity function parameters show $1\sigma$ overlap with the results of \citet{Hooper:2016rap} and the updated analysis of \citet{Amerio:2024qor}. We stress that, when assuming that the bulge MSP population follows the same GC-based luminosity function we derived here, any conclusions about the excess drawn from source counts in existing catalogs depend heavily on the assumed detection sensitivity model\cite{Dinsmore_2022,Amerio:2024qor}.

\paragraph{Advantages and limitations of our simulation-based inference approach.}
Our analysis differs from past approaches in two key aspects. First, rather than fitting only integrated GC fluxes, we condition on the \emph{full} gamma-ray spectral data of each cluster, enabling the inference to exploit information beyond a single-band luminosity summary statistic. Second, we treat the cluster-specific MSP counts as explicit variables of our simulator and infer them jointly with the luminosity function hyperparameters within the SBI framework that uses broad, agnostic priors on $N_I$ 
rather than imposing encounter-rate/optical-proxy-informed constraints. This design reduces reliance on external calibrations and allows uncertainties to propagate self-consistently through a hierarchical model without requiring an analytic likelihood for the high-dimensional spectral data.

These advantages come with corresponding caveats. The SBI posterior is conditional on the fidelity of the forward model for the cluster emission (including the common MSP spectral profile of an exponentially cut-off power law and the assumed level of diversity, the stochastic sampling of individual MSP luminosities, and any implicit assumptions about backgrounds and selection originating in the 4FGL-DR4 analysis). In addition, the inference retains an intrinsic degeneracy between source multiplicity and per-source luminosity that is only partially broken by the available spectral information, particularly for clusters near the detection threshold. Therefore, we challenged our own assumption by repeating the analysis of the baseline GC sample using merely integrated fluxes in order to derive the posterior of $(L_0,\sigma_L)$ (retaining all other features of the analysis setup including ON-OFF measurements for NGC 6624 and NGC 6626). The comparison of the obtained two-dimensional posteriors is displayed in Fig.~\ref{fig:SED-vs_intFlux} of App.~\ref{app:SED-vs-flux}. We observe that the sharpness of the posteriors in both approaches is almost identical regarding the 68\% credible region, while the 95\% credible region is slightly larger when using only integrated fluxes. The resulting best-fit parameter values are consistent with each other within 1$\sigma$. Hence, the usage of SEDs is not necessarily the dominant factor that drives the quality of the inference; it is rather the inclusion of phase-resolved GC samples.

In light of these results and the reasons given above, the literature comparison should be read as a validation of broad consistency under heterogeneous assumptions, and as motivation for future work to quantify the impact of alternative simulator components (e.g.~spectral-shape systematics and selection modeling) on the inferred luminosity function parameters.

\subsection{Compatibility with radio MSP counts and the radio-gamma connection}
\label{subsec:radio-gamma-connection}

In our forward model, $N_I$ denotes the number of MSPs that contribute to the gamma-ray emission of a given GC realisation and is therefore not expected to coincide one-to-one with the number of radio-detected MSPs. For the three clusters considered in Fig.~\ref{fig:SBI-real-NMSP}, the posterior medians imply $N_{\rm Ter5}\sim70_{-{24}}^{+{42}}$ \footnote{{consistent with \cite{Ndiyavala_2019}}}, $N_{\rm 47\,Tuc}\sim13$, and $N_{\rm NGC\,1851}\sim5$. On the radio side, the MSP census in GCs is known to be incomplete and heterogeneous: survey sensitivity, propagation effects along different lines of sight, and binary acceleration or eclipses suppress detectability in a cluster-dependent manner. Deep searches continue to expand the known populations of benchmark systems such as Terzan~5 \citep{Padmanabh:2024bsz} and NGC~1851 \citep{Ridolfi:2022gmc}, and flux and polarimetric analyses of Terzan~5 point to substantial incompleteness at the low radio-luminosity end \citep{Martsen:2022svx}. Radio counts should therefore be regarded as lower limits on the intrinsic MSP content.

Moreover, the mapping between radio and gamma-ray emission is not deterministic, even for a fixed intrinsic population. The 3PC shows a substantial overlap between radio- and gamma-loud MSPs, but also establishes the existence of radio-quiet gamma-ray MSPs \citep{Fermi-LAT:2018jvx,Fermi-LAT:2023zzt}. {Attempts to relate radio, X-ray and gamma-ray emissions through modeling of pulsars have been made, as seen in \cite{2010ApJ...714..810R,2014ApJS..213....6J,2026A&A...706A.215P, 2023ApJ...954..204K, 2014ApJ...793...97K, 2011MNRAS.412.1870P}.} Using 47~Tuc as a reference system, \citet{deMenezes:2023mzx} estimate a ratio of gamma-ray to radio beam solid angles of $\omega_\gamma/\omega_{\rm rad}\simeq 1.4\pm0.9$, while population-synthesis studies calibrated against 4FGL-DR4 require either effective gamma-ray beaming fractions of only $\sim$25--50\% or a revised luminosity scaling \citep{Tabassum:2025hzm}. Theoretical and population-synthesis work further treats radio and gamma-ray emission as coupled but geometrically distinct, rendering detectability in either band strongly geometry-driven \citep{Philippov:2019qud,Philippov:2020jxu,Sautron:2025wvb}. Consequently, $N_I$ inferred from GeV data is expected to track the number of gamma-visible and gamma-luminous MSPs rather than the total intrinsic population.

With these considerations, the inferred ordering and magnitude of $N_I$ for Terzan~5, 47~Tuc, and NGC~1851 are qualitatively consistent with these being radio-MSP-rich systems, while allowing for a gamma-loud fraction below unity. For Terzan~5, the radio-population inference of \citet{Berteaud:2024zda} yields an underlying population of order $\sim\!10^2$ MSPs (with large uncertainties). Compared with our $N_{\rm Ter5}$, this suggests that only a subset of the intrinsic MSP population dominates the GeV output, consistent with a broad intrinsic scatter in MSP gamma-ray luminosities and with non-trivial gamma-ray visibility fractions.

\section{Conclusion}
\label{sec:conclusion}

{By only taking advantage of the gamma-ray spectral emission information, in this work we constrain the gamma-ray luminosity function of MSPs in GCs.}
To this end, we developed and validated an SBI framework, {which implicitly learns a likelihood from simulated samples instead of relying on an explicitly defined one.}
A key methodological outcome is that a traditional likelihood-based multi-target analysis with nested sampling becomes computationally prohibitive already for small GC ensembles, with runtimes rising steeply as additional GCs are introduced. By contrast, the SBI approach amortizes the inference cost through NRE and therefore remains tractable when scaling to tens of targets.
On synthetic datasets for which the likelihood-based approach remains feasible, we find that SBI posteriors are consistent with their likelihood-based counterparts, and Bayesian coverage tests demonstrate that the inferred credible regions are well calibrated, supporting the reliability of the SBI pipeline when applied to real \textit{Fermi}-LAT SEDs.

Our forward model accounts for stochastic sampling of individual MSP luminosities from a shared log-normal luminosity function and propagates this discreteness into the predicted GC SEDs, thereby enabling a joint treatment of population-level luminosity function hyperparameters and (where desired) cluster-level MSP counts.
Our tests on synthetic spectra demonstrate that the inclusion of more and more GCs is highly beneficial for the inference of the GLF hyperparameters, and that it is not a single GC that drives the precision of the inference. In this respect, beating the bottleneck of computation time with our SBI framework has been key to enable us to combine multiple GCs.

In simulated data analysis, we further demonstrated that the most informative observables arise when phase-resolved ON/OFF SEDs of GCs hosting an individually detected gamma-ray MSP are available, because they partially disentangle the contribution of a single bright source from the cumulative emission of the remaining population.

Applying the method to the baseline target dataset of 36 GCs yields informative constraints on the MSP luminosity function hyperparameters, excluding a substantial fraction of the adopted prior volume. For the log-normal luminosity function, we infer
\begin{equation}
\log_{10} L_{0} = 33.3^{+0.2}_{-0.3}, \qquad
\sigma_{L} = 1.2^{+0.3}_{-0.5}
\end{equation}
(credible intervals quoted at the 68\% level), and we find that the inference largely converges once the GC sample reaches this size, with additional (fainter) targets mainly improving precision rather than shifting the preferred region. We further confirm, in the real-data application, the added value of including phase-resolved ON/OFF SEDs of GCs hosting an individually detected gamma-ray MSP. The inclusion of the full SEDs as compared to using the integrated gamma-ray fluxes of each GC is found to lead to a sharper 95\% credible region for the luminosity function parameters, which remain fully compatible at $1\sigma$ level.

Beyond luminosity function hyperparameters, we demonstrated that the same SBI framework can deliver joint constraints on cluster-specific MSP counts for selected bright GCs while maintaining broad, agnostic priors on these discrete normalizations. For Terzan~5, 47~Tucanae, and NGC~1851 the resulting marginal posteriors are unimodal and remain well confined within the prior bounds. At the same time, the inferred credible bands of the MSP counts per cluster are relatively broad, reflecting the fundamental degeneracy between source multiplicity and per-source luminosity in unresolved populations, and the limited information content of binned gamma-ray SEDs for clusters near the detection threshold.

Finally, we placed our baseline-data constraints in the context of existing findings of MSP gamma-ray luminosity functions. Despite heterogeneous target populations (Galactic field MSPs versus GC MSPs) and methodological choices across the literature, our 95\% credible region broadly overlaps with the envelope spanned by previous results. Given the present uncertainties, this comparison is best interpreted as evidence for overall consistency rather than as a discriminator among competing modelling assumptions. 

The principal advantages of our approach are (i) the use of full spectral and phase-resolved information and (ii) the absence of external, proxy-informed constraints on the expected MSP content of each cluster, with all uncertainties propagated self-consistently through the simulator.

Looking forward, the same SBI infrastructure provides a natural path to improve realism and constraining power: incorporating more phase-resolved GC SEDs as they become available; extending the forward model to include spectral-shape systematics and selection effects; and relaxing the assumption of a single universal luminosity function through physically motivated hierarchical extensions (e.g., efficiency and evolutionary dependencies). Our log-normal prescription is phenomenological and compresses several layers of pulsar physics into two effective parameters. Part of the inferred width $\sigma_L$ likely reflects the broad dispersion of gamma-ray conversion efficiencies $\eta_\gamma = L_\gamma/\dot{E}$, which for \emph{Fermi}-LAT pulsars is dominated by an intrinsic scatter of order one dex once beaming and distance uncertainties are accounted for \citep{Iniguez-Pascual:2024jal, Iniguez-Pascual:2025kgt, Iniguez-Pascual:2025eqg}. In addition, spin-down evolution progressively depletes the bright end of an ageing MSP population on $\mathcal{O}(0.1$--$1)\,\mathrm{Gyr}$ timescales \citep{Hooper:2016rap}, so that clusters with different MSP recycling histories may deviate from a strictly universal luminosity function even if the underlying microphysics is identical. Treating the efficiency, beaming factor and recycling-age distribution as explicit latent variables would allow a more direct interpretation of the inferred GLF parameters in terms of MSP energetics and evolutionary state, at the cost of new degeneracies that will require additional spectral or multi-wavelength constraints. {Specific models relating the gamma-ray luminosity of MSPs and their $\dot{E}$ may thus be considered to further constrain the GLF parameters \citep{2026ApJ...999....5K}}. {The same work and its predecessor \cite{2019ApJ...883L...4K} also establish spectral-luminosity correlations that could be employed in future works.}

These developments will be essential for turning GC gamma-ray spectra into more precise probes of MSP population physics.


\begin{acknowledgements}
The authors gratefully acknowledge the HPC RIVR consortium \href{https://www.hpc-rivr.si}{(www.hpc-rivr.si)} and EuroHPC JU \href{https://eurohpc-ju.europa.eu/}{(eurohpc-ju.europa.eu)} for funding this research by providing computing resources of the HPC system Vega at the Institute of Information Science \href{https://www.izum.si/en/home/}{(www.izum.si)}.

The work of FC and CE is supported by the European Union through the grant ``UNDARK'' of the Widening participation and spreading excellence programme (project number 101159929). This publication has received funding from the European Union’s Horizon Europe research and innovation program under the Marie Sk\l odowska-Curie COFUND Postdoctoral Programme grant agreement No.101081355-SMASH and from the Republic of Slovenia and the European Union from the European Regional Development Fund. 
SM acknowledges the support of the French Agence Nationale de la Recherche (ANR) under the grant ANR-24-CPJ1-0121-01, and support of the European Union's Horizon Europe research and innovation program for support under the Marie Sklodowska-Curie Action HE MSCA PF–2021,  grant agreement No.10106280, project \textit{VerSi}. The work of FC is supported by the Agence Nationale de la
Recherche through the project ANR-25-CE31-6485. 
\textbf{Disclaimer}: Co-funded by the European Union. Views and opinions expressed are, however, those of the authors only and do not necessarily reflect those of the European Union or European Research Executive Agency. Neither the European Union nor the granting authority can be held responsible for them.
{The {\it Fermi}-LAT Collaboration acknowledges generous ongoing support from a number of agencies and institutes that have supported both the development and the operation of the LAT as well as scientific data analysis. These include the National Aeronautics and Space Administration and the Department of Energy in the United States, the Commissariat à l'Energie Atomique and the Centre National de la Recherche Scientifique / Institut National de Physique Nucléaire et de Physique des Particules in France, the Agenzia Spaziale Italiana and the Istituto Nazionale di Fisica Nucleare in Italy, the Ministry of Education, Culture, Sports, Science and Technology (MEXT), High Energy Accelerator Research Organization (KEK) and Japan Aerospace Exploration Agency (JAXA) in Japan, and the K. A. Wallenberg Foundation, the Swedish Research Council and the Swedish National Space Board in Sweden.
Additional support for science analysis during the operations phase is gratefully acknowledged from the Istituto Nazionale di Astrofisica in Italy and the Centre National d'Etudes Spatiales in France. This work performed in part under DOE Contract DE- AC02-76SF00515.}
\end{acknowledgements}

\bibliographystyle{aa}
\bibliography{refsdraft}

\appendix

\section{\textit{Fermi}-LAT data selection and preparation}
\label{sec:fermi-data}
In this appendix, we indicate the settings used in our \textit{Fermi}-LAT data acquisition.\\
For each GC, we select 17 years of \textit{Fermi}-LAT events (\texttt{evclass=128}) spanning the time interval between 2008 August 4, 15:43:36.0 (UTC) and 2025 December 14, 15:42:56.0 (UTC), covering energies from $0.05$ to $100~\mathrm{GeV}$. We start with all the events within a search radius of $14^\circ$ around each GC, while we analyse events in a square region of interest (RoI) of $17^\circ \times 17^\circ$ with the same center. Notice that the search radius is chosen to be larger than the RoI width to correctly include photons from extended sources which may partially lie outside the RoI. 
The spatial bin size is set to $0.1^\circ$. Data are filtered using the expression \texttt{DATA\_QUAL==1 \&\& LAT\_CONFIG==1} to remove events flagged as poor quality {(\texttt{DATA\_QUAL==0}) or not suitable for pulsar timing (\texttt{DATA\_QUAL==2})}. 
For the instrument response function, we adopt the Pass~8 Version~3 release (\texttt{P8R3\_SOURCE\_V3}, see \cite{2013arXiv1303.3514A,2018arXiv181011394B}).\\
For energies between 50 MeV and 500 MeV, we set \textit{evtype} to 48, while for energies between 500 MeV and 1 GeV it is set to 56 and for energies between 1 GeV to 100 GeV to 3. {The \textit{evtype} represents the event type used in \textit{gtselect} and it incorporates both the partition in PSF and in the energy dispersion type. For every subselection with a specific \textit{evtype} $n_i$, the \textit{evtype} of the full selection is given as $\sum_i n_i$. Hence, a value of 48 represents a selection of PSF3 and PSF2, while 56 represents PSF{1}+PSF2+PSF3. Finally, an evtype of 3 indicates FRONT+BACK events \footnote{The meaning of the PSF partitions and an explanation of the \textit{evtype} and \textit{evclass} can be found in \href{https://fermi.gsfc.nasa.gov/ssc/data/analysis/documentation/Cicerone/Cicerone_Data/LAT_DP.html}{Cicerone: Data - LAT Data Products} and \href{https://fermi.gsfc.nasa.gov/ssc/data/analysis/documentation/Pass8_usage.html}{Pass 8 Data Usage}.}.}
To these energy ranges, we apply a zenith cut of $z_{max}=80^\circ,100^\circ,105^\circ$ respectively. The configuration files were produced using \textsc{EasyFermi v.2.0.20} \citep{EasyFermi}, with only minor adjustments to account for the phase resolved analysis described here.

\section{Spectral Energy Distribution for NGC 6624 and NGC 6626}
\label{app:sed-on-off}

\begin{figure*}[t!]
    \centering
    \includegraphics[width=0.49\linewidth]{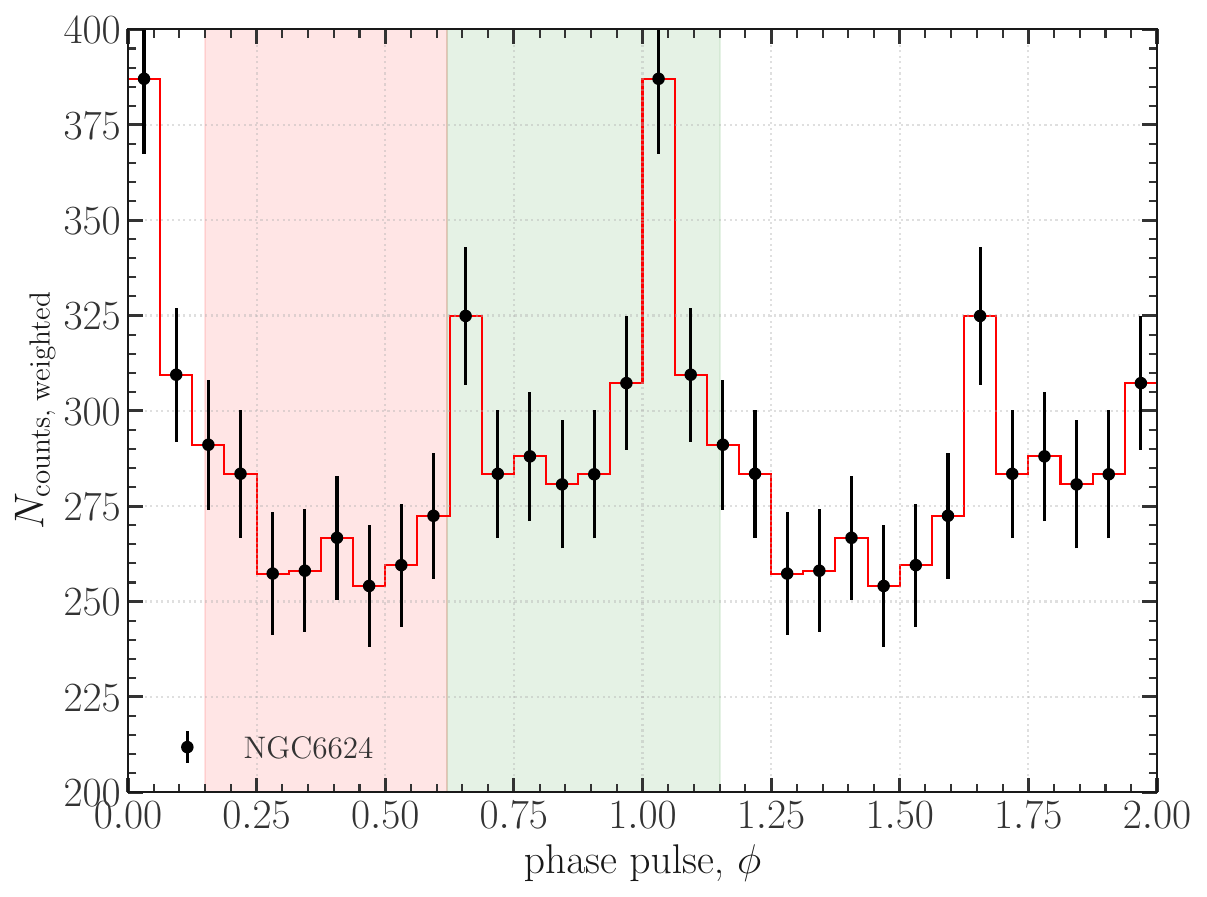}\hfill
    \includegraphics[width=0.49\linewidth]{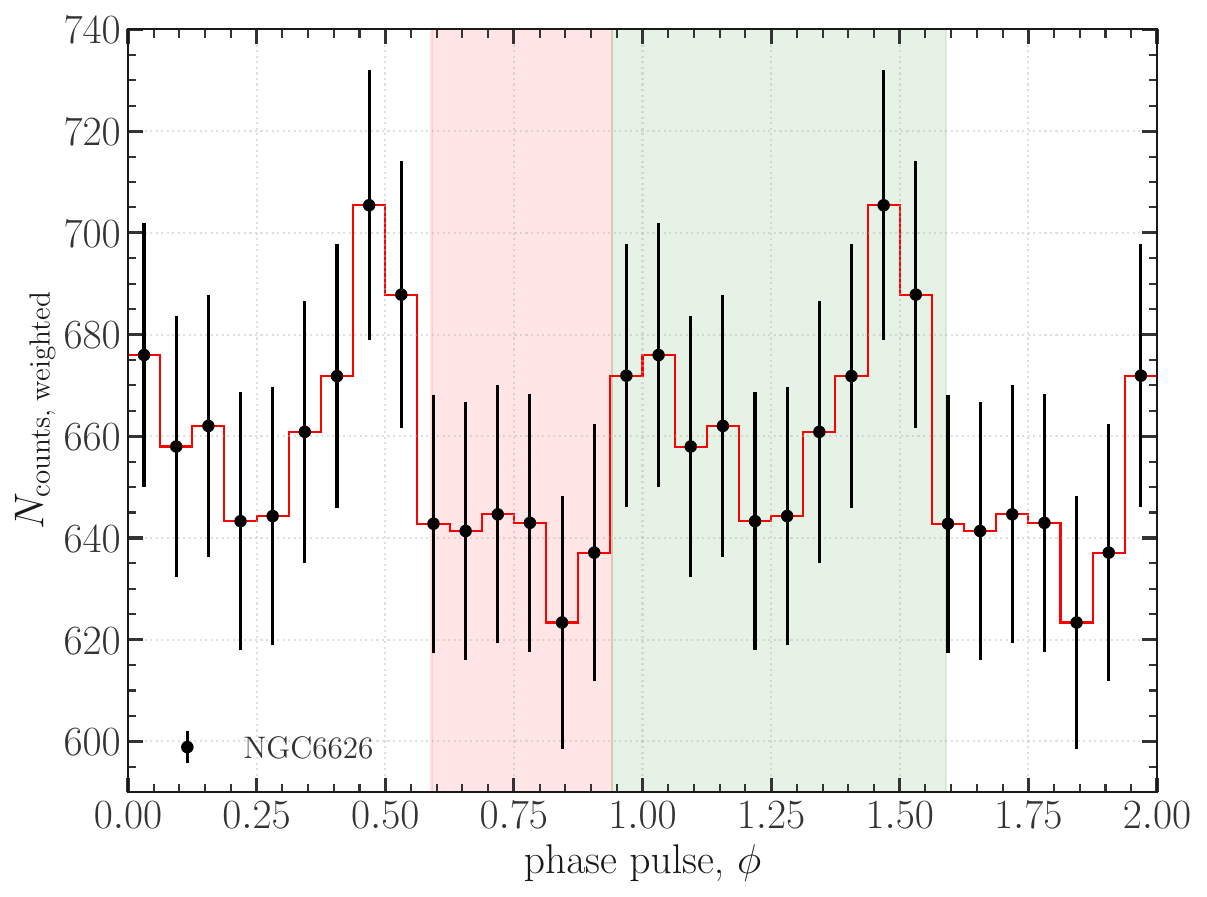}
    \caption{Weighted phaseograms for the two GCs containing only 1 MSP, but having a significant OFF-phase emission. We repeat the phaseograms twice (from 0.0 to 2.0), so that we can define continuous intervals for both the ON- and OFF-phases. The shaded green (red) regions represent the ON(OFF)-phase intervals, also shown in Table \ref{tab:onofftable}.\textbf{Left:} NGC 6624. \textbf{Right:} NGC 6626.}
    \label{fig:phaseograms}
\end{figure*}

\begin{figure*}[t!]
    \centering
    \includegraphics[width=0.49\linewidth]{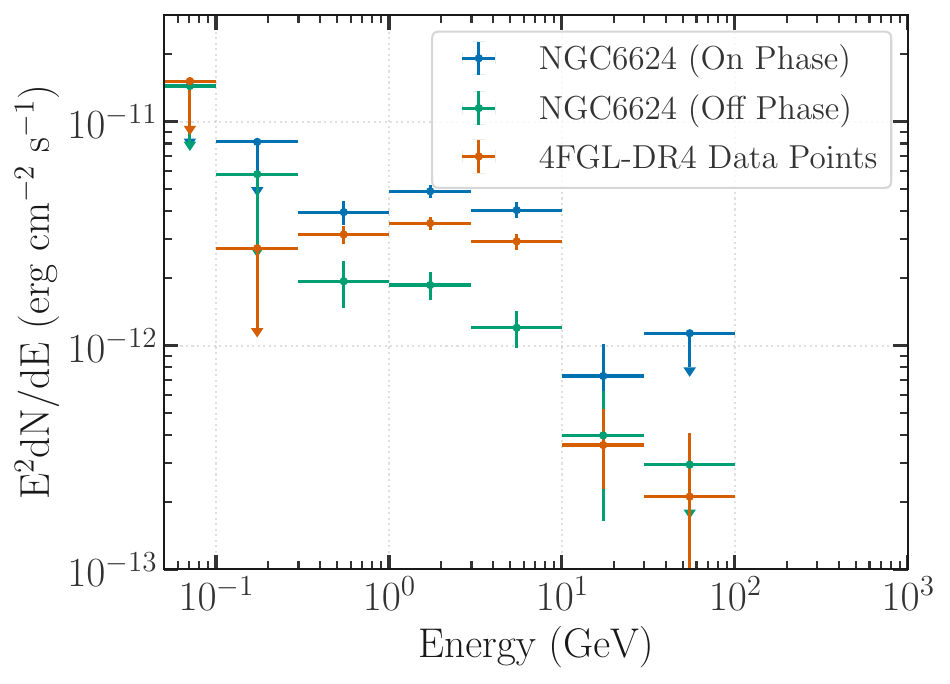}\hfill
    \includegraphics[width=0.49\linewidth]{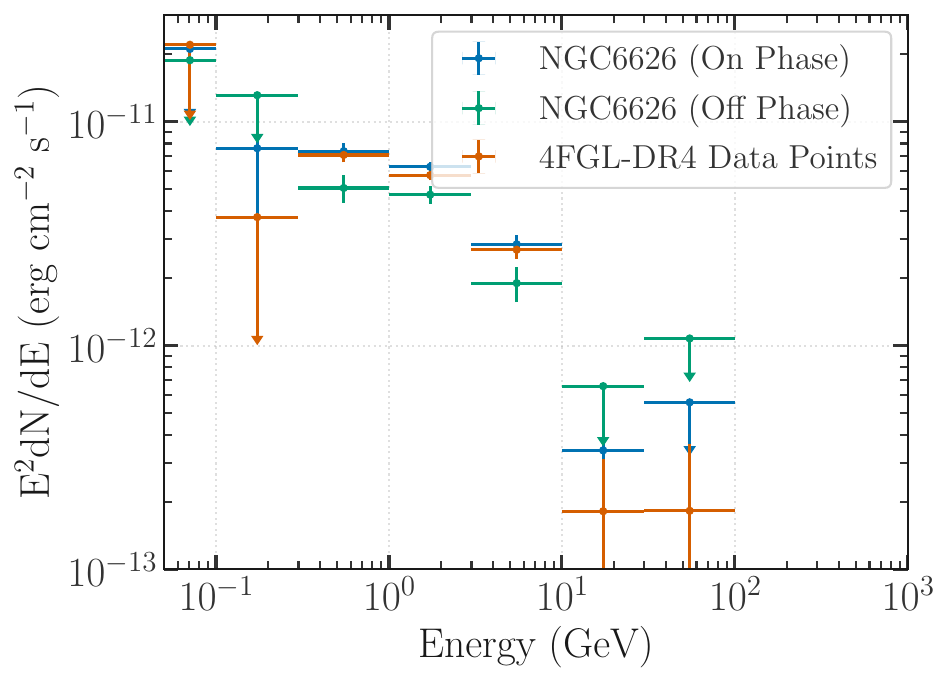}
    \caption{Comparison between our spectra both ON- (blue points) and OFF-phase (green points) with 4FGL spectra (orange points). We require for any point that $TS\geq4$ (threshold value for a $2\sigma$ detection) and show $1\sigma$ uncertainties. If $TS<4$, we find upper limits at a $95\%$ confidence level. \textbf{Left:} NGC 6624. \textbf{Right:} NGC 6626.}
    \label{fig:SEDs_1MSP}
\end{figure*}

As stated in section \ref{sec:gc-sed}, we derive the SEDs for two GCs, NGC6624 and NGC6626. We first determine their phaseograms, from which we obtain the ON and OFF phases detailed in Table \ref{tab:onofftable} by using the Bayesian Block algorithm (see refs. \cite{Jackson_2005,Scargle_2013}) on the unweighted phaseograms following the approach described in the 2PC \citep{2013ApJS..208...17A}. We show the weighted phaseograms in Figure \ref{fig:phaseograms}, where in green we represent the ON-phase region and in red the OFF-phase ones. The phaseograms are repeated twice, transposing the pulsar's phases to an interval between 0 to 2 instead of from 0 to 1, so that we can define continuous ON and OFF-phase regions. For both NGC 6624 and NGC 6626 we end up having similar ON-OFF Phase intervals as \cite{Zhang:2023gwr}.
\\
We follow section \ref{sec:gc-sed} to obtain their ON-phase and OFF-phase SEDs, which we show in Figure \ref{fig:SEDs_1MSP}, together with a comparison to the 4FGL-DR4 results (from which we adopt the energy binning). For both our analysis and the 4FGL-DR4 points we show detections at $2\sigma$ for each energy bin. For the test statistic ($TS$), this translates to requiring $TS\geq4$. If $TS<4$, we find upper limits at a $95\%$ confidence level.\\
As can be seen, a significant OFF-phase emission is expected from both NGC 6624 and NGC 6626 as reported in \citep{Zhang:2023gwr}. 

\section{Supplementary figures related to simulated data}
\label{app:large-figures}

\subsection{Matching \textit{Fermi}-LAT spectra with synthetic millisecond-pulsar populations}
\label{app:synthetic-preparation}

\begin{figure}[t!]
    \centering
    \includegraphics[width=\columnwidth]{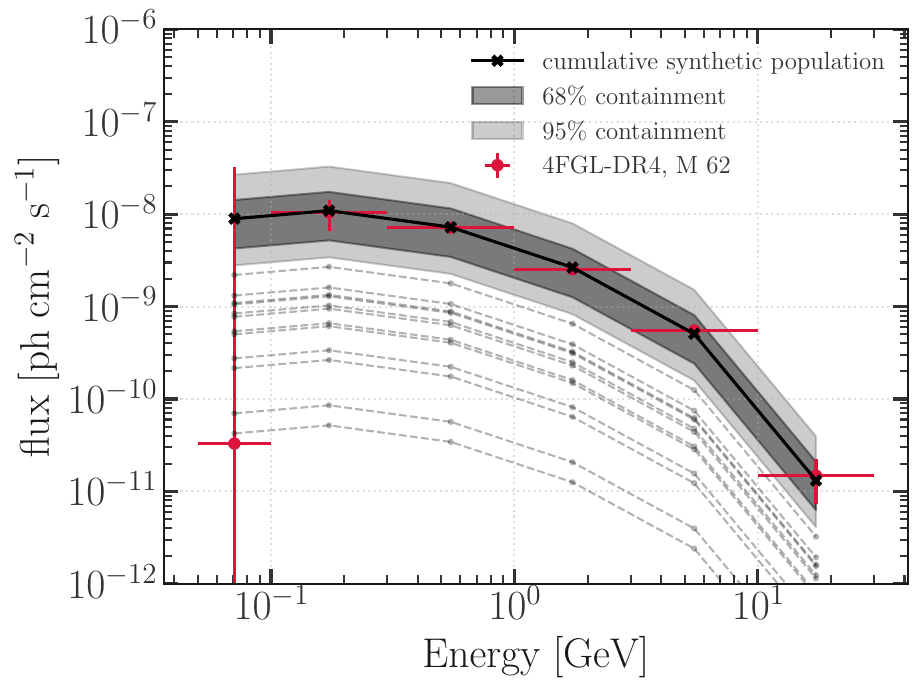}
    \caption{Preparation of a synthetic gamma-ray SED for M 62 based on the observed gamma-ray flux reported in 4FGL-DR4 (red points and error bars). The shaded bands in grey represent the 68\% and 95\% containment region obtained from 250 MSP populations simulated from a gamma-ray luminosity function with $L^s_0=3\times10^{33}$ erg/s and $\sigma^s_L=1.6$ for $N^s_{\rm M 62}=12$ individual MSPs. The individual MSPs' SEDs in this realization are displayed with dashed grey lines, while their cumulative gamma-ray SED is shown as a solid black line.}
    \label{fig:synthetic-sed-example}
\end{figure}

In Fig.~\ref{fig:synthetic-sed-example}, we provide a concrete example based on M 62 of how we derived synthetic gamma-ray SEDs for the GCs listed in Tab.~\ref{tab:synthetic-SEDs}. As mentioned in the main text, we first fix the hyperparameters of the MSP luminosity function to be $L^s_0=3\times10^{33}$ erg/s and $\sigma^s_L=1.6$. Then, we use the gamma-ray SED of the observed GC in question, here M 62, which is displayed with red data points. To obtain a suitable number of MSPs hosted in the synthetic analog of M 62, we generate 250 realizations of its MSP population by successively increasing $N^s_{\rm M 62}$ until the 68\% containment band for the cumulative SED of all synthetic MSPs as obtained from these 250 realizations fully encompasses the real data points within their $1\sigma$ uncertainty. This containment band is shown in grey while the cumulative SED for the one selected realization is given as a solid black line. To this end, we assume that all simulated MSPs follow the {representative} PLEC4 spectral profile obtained from the MSPs listed in 3PC (see Sec.~\ref{sec:generate-synthetic} for details). In addition, we plot all individual MSP SEDs with dashed grey lines. For the actual inference task, we pick one realization among the set of generated SEDs whose cumulative SED matches the real LAT data points within $1\sigma$ (as shown here in the example).

\section{Network architecture and hyperparameter definition}
\label{sec:sbi-nn}

As noted earlier, inferring model parameters with \texttt{swyft} requires the definition of neural network architectures to perform the binary classification task and, if the user wishes to do so, to learn a summary statistic compressing the simulator's output into a lower-dimensional representation. Both neural networks impact the robustness and quality of the inference; they need to be chosen such that the summary statistic is flexible enough to learn the correlations between model parameters. \texttt{swyft} itself leverages the ML library \texttt{pytorch-lightning} \citep{PyTorch_Lightning_2019} to construct, train and optimize the involved neural networks, which we hence adopt for all SBI-related steps. 

\noindent\textit{Neural network for summary statistic.} For this task, we utilize a simple feed-forward deep neural network with two hidden layers. The simulator generates a data vector of a certain length depending on the selected set of GCs. A linear layer maps the input into an array of size 256. The following two hidden layers maintain the size of the array while the last layer performs a feature reduction and outputs an array with 64 features. After each layer -- except for the final one -- we employ a ``LeakyReLU'' activation function. The final number of features is large enough to learn the correlations among the model parameters for a reasonable number of stacked GCs. We will explicitly state later in the text whenever we modify these baseline definitions.

\noindent\textit{Neural network for binary classification task.} We exploit \texttt{swyft}'s built-in residual neural network \citep{He:2015wrn} to this end. We use three residual blocks with 128 hidden features each. On top of that, we allow for dropout layers with a 20\% dropout rate. 

A crucial feature of NRE is its ability to single out parameters of interest for which low-dimensional marginal posteriors are derived. There is no need to obtain the full $D$-dimensional joint posterior of all parameters of the model. This ability saves a substantial amount of computation time. In this work, we focus on deriving one- and two-dimensional marginal posterior distributions for all parameters. In some cases, we even marginalize over the number of MSPs per GC $N_I$ and only infer the tuple $(L_0, \sigma_L)$. For the one- and two-dimensional marginals, we need to build an associated neural network, which follows the layout stated above.

\noindent\textit{Training and hyperparameters.} We train the full \texttt{swyft} inference pipeline with the following baseline specifications: Per GC composition, we generate a sample of $10^6$ data vectors for which 80\% serve as training data and 20\% as validation data. They are loaded in batches of 128 samples. The training lasts a maximum of 250 epochs with an initial learning rate of $10^{-4}$. We implement an early stopping patience of 20 based on the lowest validation loss averaged over an epoch. Additionally, we utilize a learning rate scheduler that reduces the learning rate by a factor of 0.1 whenever the validation loss does not decrease for eight epochs. The network training leverages the `AdamW' optimizer \citep{2017arXiv171105101L} (with default parameters of \texttt{pytorch}). To infer the parameters of a target data vector, we draw $10^5$ prior samples. 

It is a special feature of \texttt{swyft} to allow for noise resampling per epoch. Since the simulator structure saves all data products from the drawn parameters to the final data vector, it is possible to redraw $\tilde{F}_{k, I}$ from $F_{k, I}$ in each epoch. Doing so increases the variability of the training dataset and lowers the risk of overfitting the training sample. In fact, it also helps the neural network to learn the noise itself much better than without resampling. Therefore, we activate this option for all cases.

\section{Further simulation-based inference result figures on the set of synthetic globular clusters}
\label{app:8GC-synthetic}

\begin{figure*}
    \centering
    \includegraphics[width=0.99\linewidth]{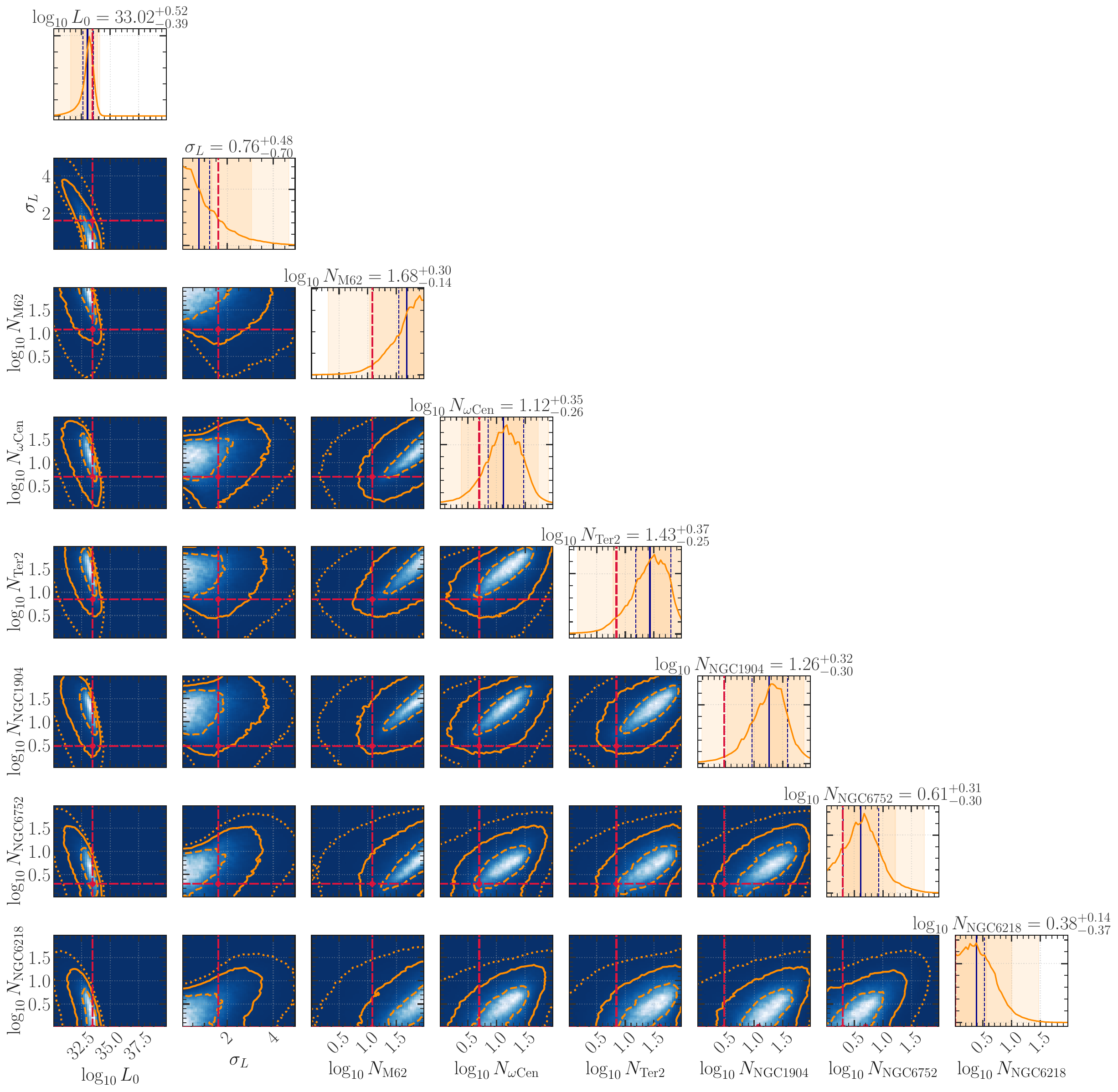}
    \caption{Same as Fig.~\ref{fig:results-dynesty} using the SBI approach to derive the two-dimensional marginal posteriors exploiting synthetically generated data of the six GCs specified in Sec.~\ref{sec:generate-synthetic} that do not feature a detected gamma-ray MSP.}
    \label{fig:SBI-6GC-noConstraints}
\end{figure*}

In Fig.~\ref{fig:SBI-6GC-noConstraints}, we display the results of SBI on the set of six synthetic GCs that do not feature a detected MSP; i.e., excluding NGC 6341, NGC 6652 and NGC 6624. In comparison to Fig.~\ref{fig:SBI-vs-dynesty}, these posteriors demonstrate that the addition of at least one GC with a constraint on the detected number of MSPs significantly improves the inference quality regarding essentially all model parameters. We observe that without any additional information on the number of MSPs in a given GC, the posteriors of $N_I$ may be biased systematically; here, towards larger values because $\sigma_L$ runs against the lower boundary of its prior range.

\begin{figure*}
    \centering
    \includegraphics[width=0.99\linewidth]{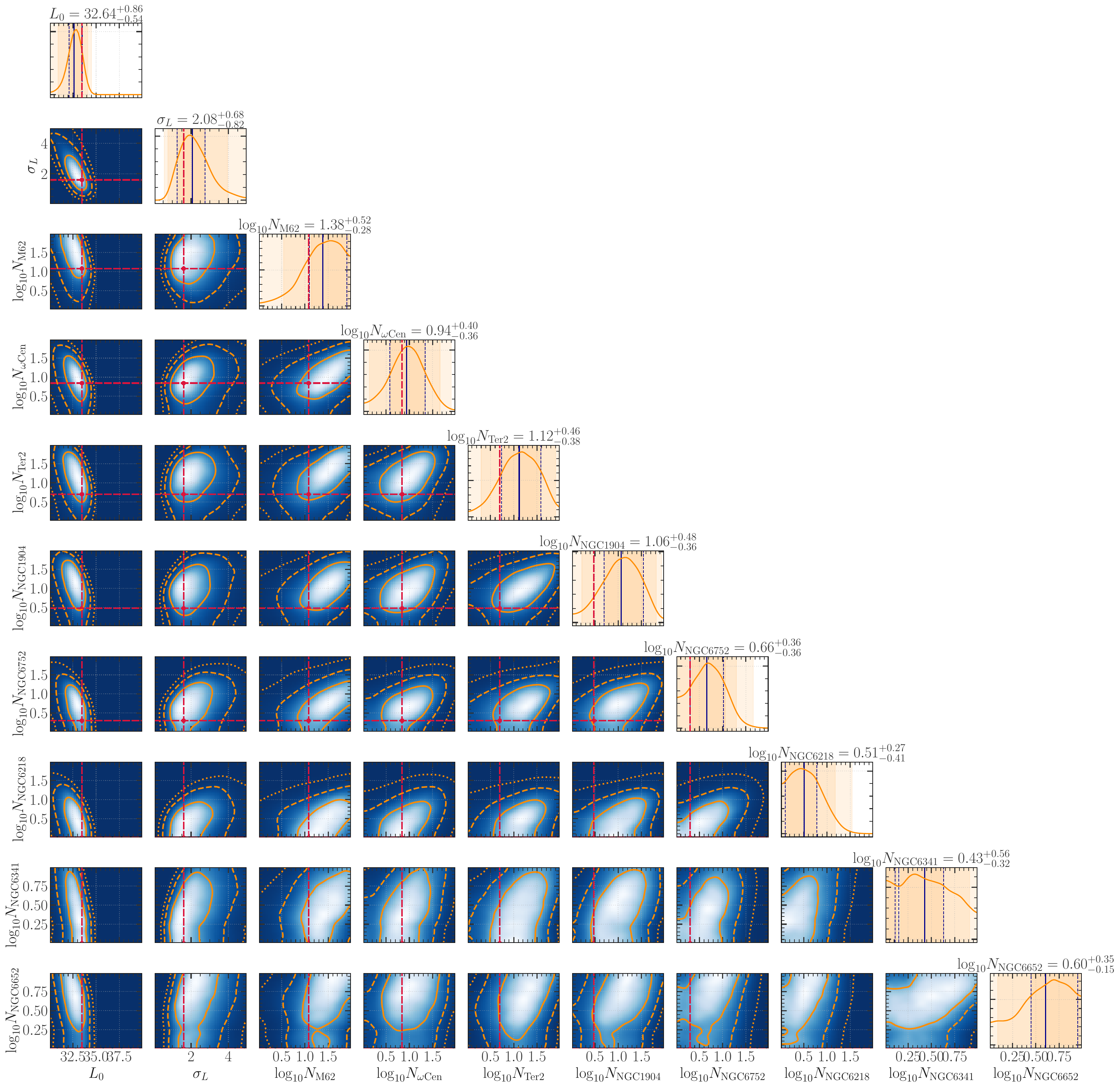}
    \caption{Same as Fig.~\ref{fig:results-dynesty} using the SBI approach to derive the two-dimensional marginal posteriors exploiting synthetically generated data of the eight GCs specified in Sec.~\ref{sec:generate-synthetic}.}
    \label{fig:SBI-allGCs-noONOFF}
\end{figure*}

In Fig.~\ref{fig:SBI-allGCs-noONOFF}, we report the results of our SBI on the set of all synthetic GCs SEDs, which do not feature significant OFF-phase emission of a detected MSP; i.e.~excluding NGC 6624. In comparison to Fig.~\ref{fig:SBI-vs-dynesty}, these posteriors demonstrate that adding more GC SEDs helps to narrow down the MSP gamma-ray luminosity function parameters.

\begin{figure*}
    \centering
    \includegraphics[width=0.99\linewidth]{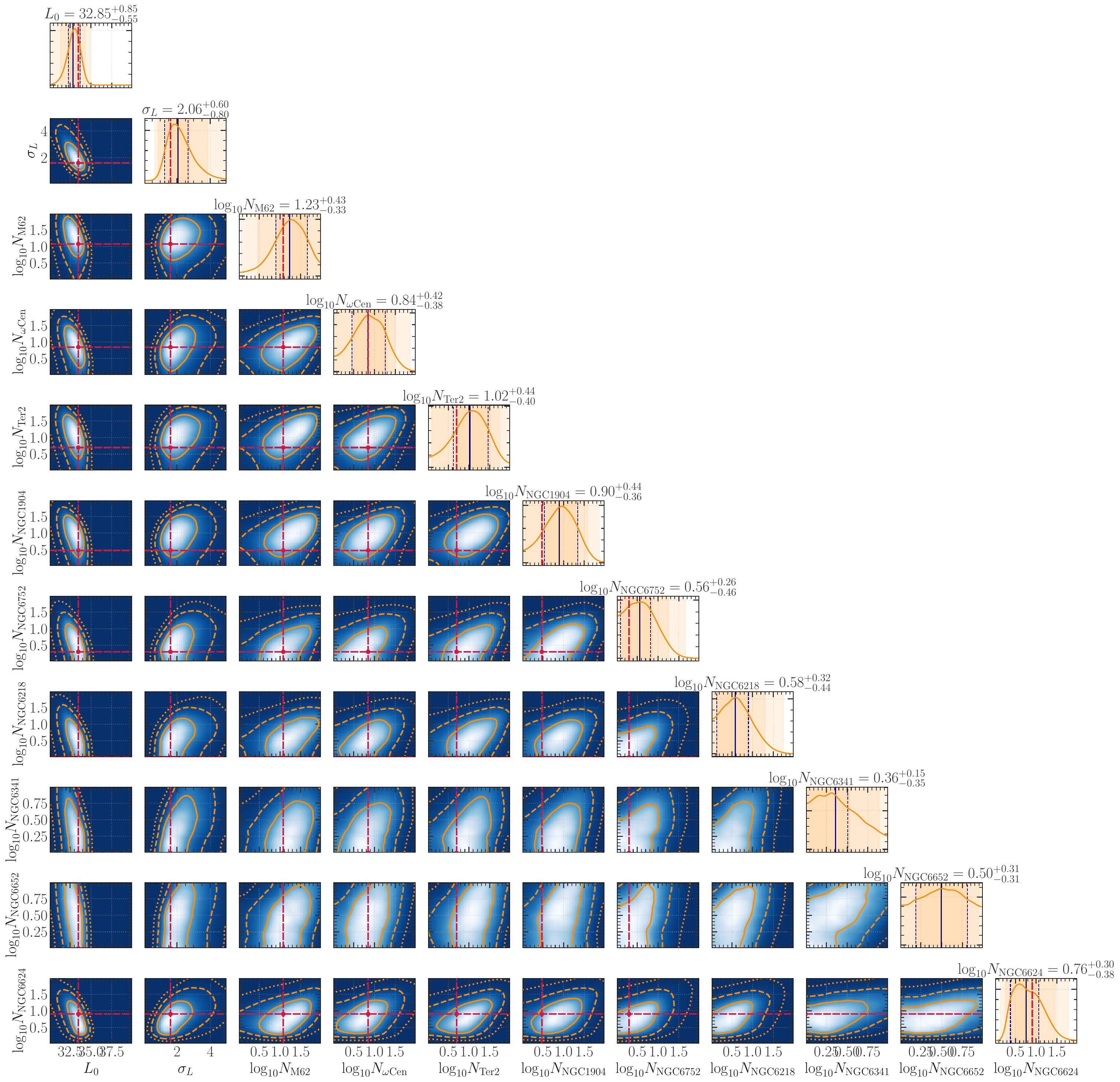}
    \caption{Same as Fig.~\ref{fig:SBI-allGCs-noONOFF} using the SBI approach to derive the two-dimensional marginal posteriors exploiting synthetically generated data of the eight GCs specified in Sec.~\ref{sec:generate-synthetic} and adding NGC 6624 for which we generated two SEDs -- ON-phase emission related to a single detected gamma-ray MSP and the remaining OFF-phase emission of the cumulative MSP population remainder.}
    \label{fig:SBI-allGCs-wONOFF}
\end{figure*}

In Fig.~\ref{fig:SBI-allGCs-wONOFF}, we show the inference on the full set of synthetic GC SEDs, this time, including one cluster featuring significant ON- and OFF-phase emission. 

\section{Calibration and validation of the simulation-based inference analysis}
\label{app:sbi-validation}

\subsection{Posterior calibration}
\label{app:sbi-coverage}

One of the crucial questions within the framework of SBI methods is the question of the quality of the obtained posterior distributions. Generally speaking, to quantify the precision of the posteriors, that is, whether or not their shape is the best outcome one can achieve given the provided information from the target dataset, an explicit likelihood function is required from which one can run explicit Monte Carlo Markov Chains. Then, the results of both methods can be compared. In our case, we have access to the likelihood function, and we verified in Sec.~\ref{sec:sbi-results-dynesty} that likelihood-based and SBI methods produce compatible posteriors. 

However, SBI is also a powerful tool when an explicit likelihood function cannot be efficiently sampled or when it does not even exist in a closed form. In such cases, the precision question is harder to answer, but it is nonetheless possible to verify the calibration of the posteriors obtained with an SBI method via so-called coverage tests. This calibration check is nothing but a consistency check of the derived posterior credible regions, answering whether or not they correctly cover the inferred parameters on average across a large number of observations. In practice, we conduct the following coverage test: Using our GC gamma-ray simulator, we generate 1000 mock observations of our selected composition of GC SEDs. We infer, with the network trained on this particular case, the posterior distributions for each mock observation, and count how many times the credible intervals of $x \in \left[0, 100\right]$ \% encompass the true value. This yields the empirical coverage for a particular training round. Nominally, if the posteriors are perfectly calibrated, the x\% credible interval must contain the true parameter value in x\% of the cases.

\begin{figure}
    \centering
    \includegraphics[width=0.9\columnwidth]{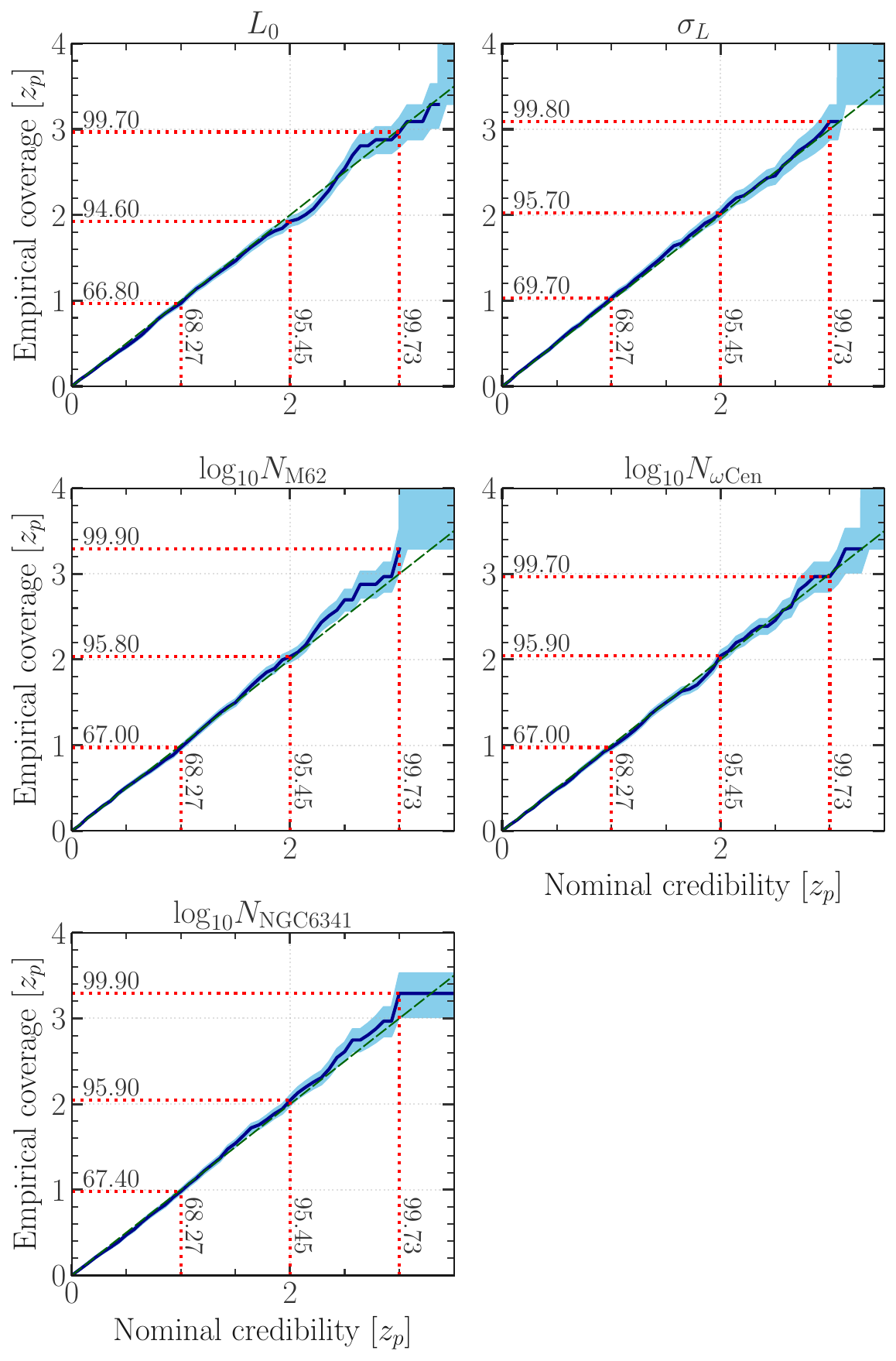}
    \caption{Coverage results for the one-dimensional marginal posterior distributions shown in Fig.~\ref{fig:SBI-vs-dynesty}.}
    \label{fig:coverage-3GC}
\end{figure}

\begin{figure}
    \centering
    \includegraphics[width=0.9\columnwidth]{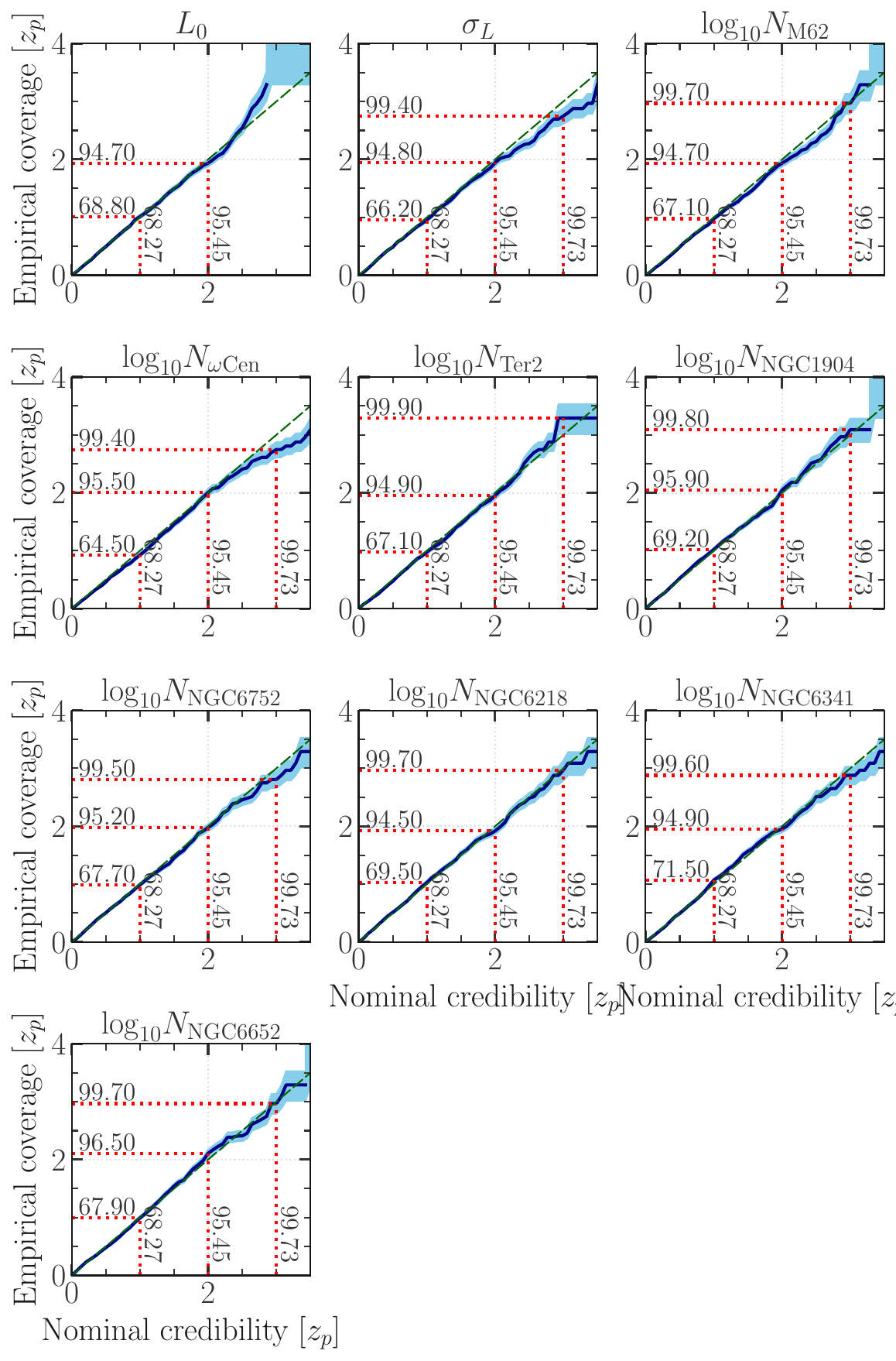}
    \caption{Coverage results for the one-dimensional marginal posterior distributions shown in the right panel of Fig.~\ref{fig:SBI-allGCs-noONOFF}.}
    \label{fig:coverage-8GC}
\end{figure}

\begin{figure}
    \centering
    \includegraphics[width=0.9\columnwidth]{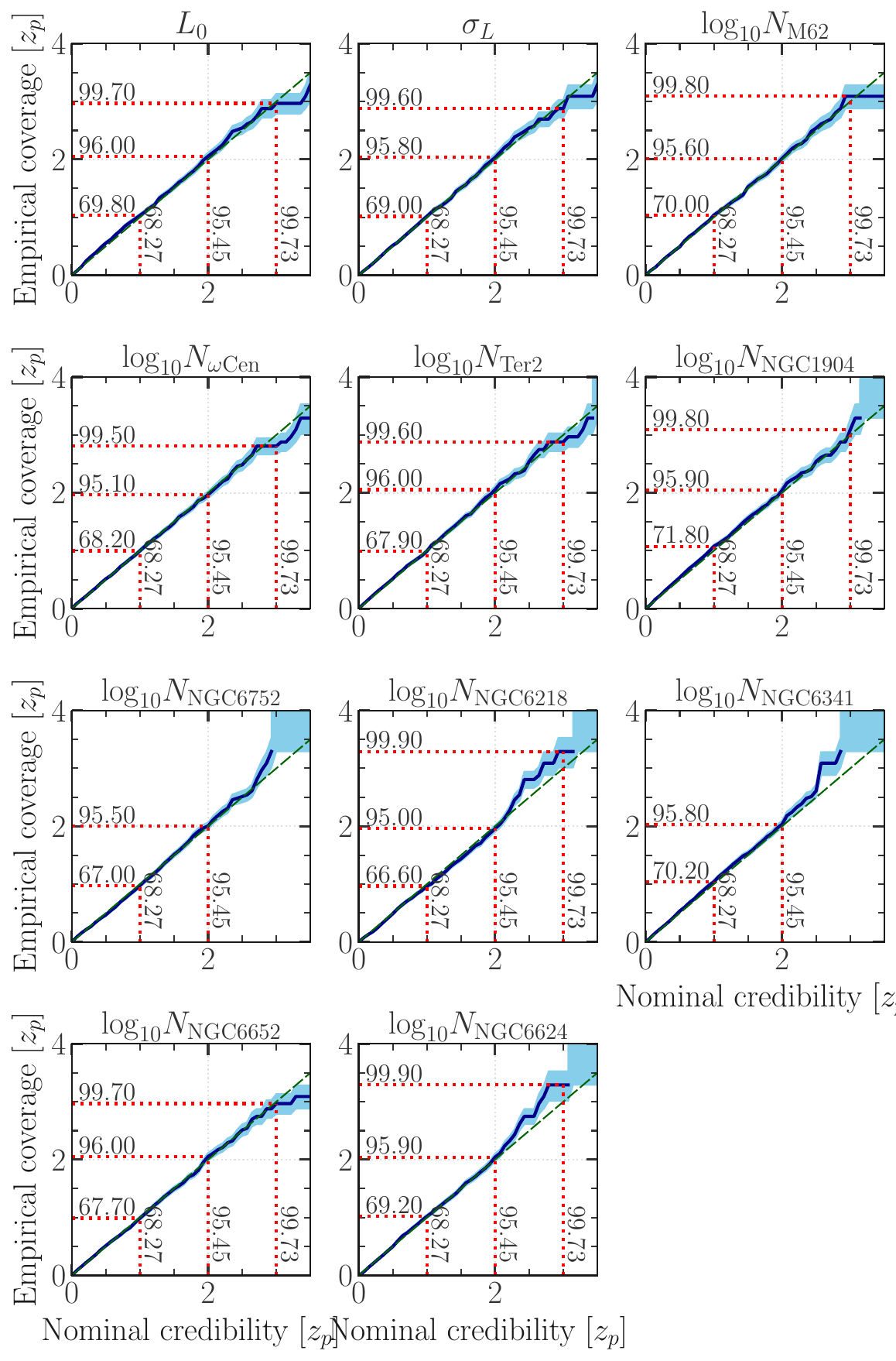}
    \caption{Coverage results for the one-dimensional marginal posterior distributions shown in the right panel of Fig.~\ref{fig:SBI-allGCs-wONOFF}.}
    \label{fig:coverage-9GC}
\end{figure}

We plot the coverage derived via this method in Figs.~\ref{fig:coverage-3GC} to \ref{fig:coverage-9GC}. The nominal credibility is plotted on the horizontal axis, while the empirical coverage is plotted on the vertical axis. Perfect calibration corresponds to a line at $45^{\circ}$ denoted as a dashed green line. The derived coverage is shown as a solid blue line, while its $1\sigma$ uncertainty is displayed as a blue shaded band. An empirical coverage above the green line indicates a conservative posterior estimate, whereas a coverage below this line implies overconfident posteriors. In fact, almost all of the obtained coverage plots indicate that our obtained posteriors are well-calibrated and only marginally overconfident regarding the $3\sigma$ credible bands in some single cases. We stress, however, that calibration is only a necessary condition for the posteriors to be a valid estimate of the true posteriors.


\subsection{Impact of full spectral energy density compared to integrated globular cluster fluxes}
\label{app:SED-vs-flux}

Throughout the main text, we worked under the assumption that the GC gamma-ray data are represented via their respective SED (as reported in 4FGL-DR4). All results in this work have been derived under this assumption. Here, we challenge this fundamental decision and repeat the analysis of our baseline GC sample merely leveraging the energy flux $\int E\,\mathrm{d}N/\mathrm{d}E\,\mathrm{d}E$ (100 MeV to 100 GeV) of each GC. The gamma-ray simulator can easily accommodate this data representation choice by integrating the spectrum per drawn MSP. In addition, our setup regarding phase-resolved GCs and the implementation of GCs with the constraint that one MSP is responsible for the entire measured emission translate naturally without any change.

\begin{figure}[h!]
    \centering
    \includegraphics[width=0.9\columnwidth]{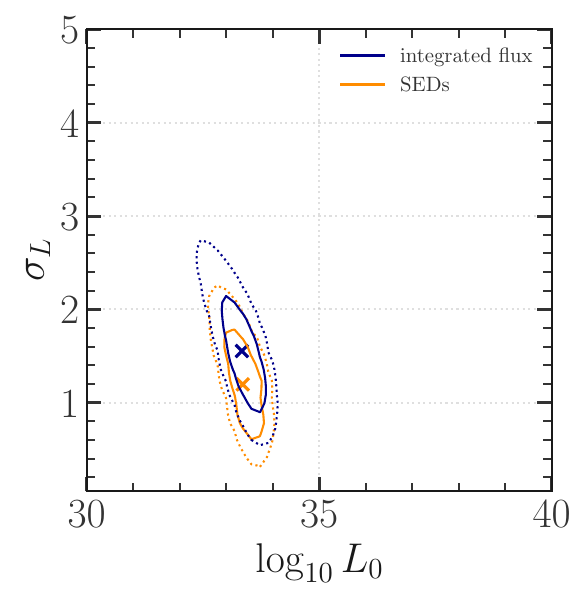}
    \caption{Comparison of the marginal two-dimensional posterior for $(L_0, \sigma_L)$ derived from the baseline GC sample using either full gamma-ray SEDs (orange) or the integrated flux per cluster (blue). The best-fit parameter tuple is denoted with a cross in both cases. {The solid/dotted lines refer to the $2\sigma$/$3\sigma$ credible intervals of the marginal two-dimensional posterior.}}
    \label{fig:SED-vs_intFlux}
\end{figure}

In Fig.~\ref{fig:SED-vs_intFlux} we show the obtained result when marginalizing over all GC MSP number counts $N_I$, i.e.~the marginal $(L_0, \sigma_L)$ posterior. There are two observations: First, the 95\% credible region of the integrated flux posterior is larger than that of the one based on SEDs. Hence, the SEDs yield a certain level of additional information that leads to sharper posteriors. Second, while the inferred best-fit values of $L_0$ perfectly align, the respective value of $\sigma_L$ differs between the two cases. We find $\sigma_L= 1.5^{+0.5}_{-0.4}$ when performing the analysis with integrated fluxes, higher than the value yielded by the same analysis with SEDs. Nonetheless, both results are compatible at the $1\sigma$ level. We can only speculate about the mechanism behind this shifted value for $\sigma_L$. A larger width of the luminosity function at the same pivot luminosity implies that it is more likely to generate brighter MSPs on average. Therefore, the observed gamma-ray emission of a GC is caused by a smaller number of MSPs (on average). Due to the variability of the intrinsic MSP gamma-ray spectra, the SED could be dominated by the bright ones. The integrated flux approach is blind to such effects, whereas the SED version of the analysis can detect and penalize cases where one would rather need the cumulative spectra of many sources rather than a few dominant MSPs.

\subsection{Prior sensitivity of the inferred parameters}
\label{app:prior-impact}

In the main text, we have shown the results for the inferred number of MSPs in selected GCs; see Fig.~\ref{fig:SBI-real-NMSP}. These posteriors have been derived with our baseline assumption of a log-uniform prior for the number of MSPs in each respective GC. To better interpret these numbers, it is crucial to test the stability of the posteriors under changes in the prior distributions, as the Bayesian approach hinges on prior knowledge. In what follows, we challenge our baseline assumption by assuming a uniform prior distribution for all $N_I$ while retaining their respective ranges in which we vary the number of MSPs per GC. 

\begin{figure}
    \centering
    \includegraphics[width=0.9\columnwidth]{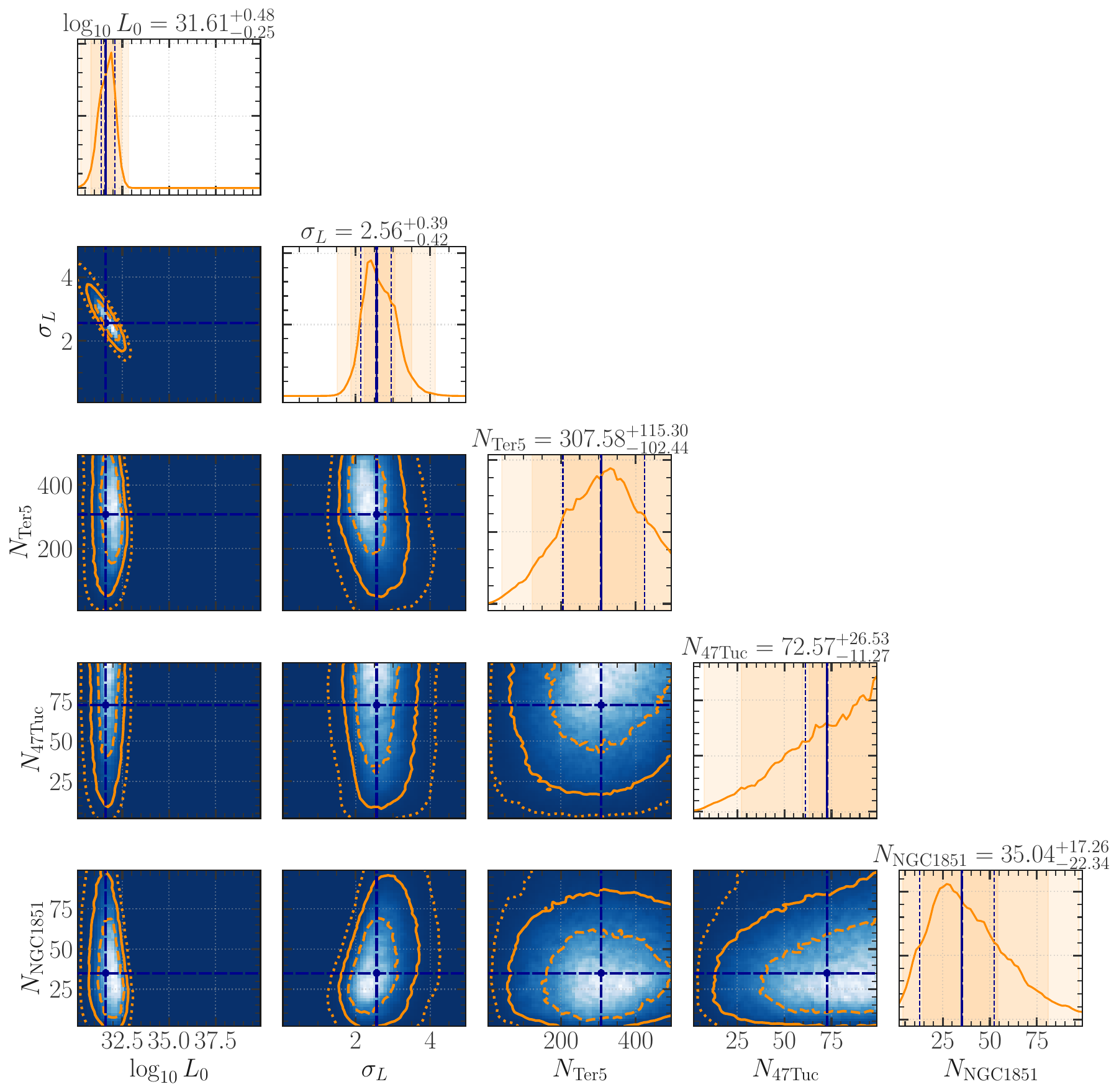}
    \caption{Same as Fig.~\ref{fig:SBI-real-NMSP} using a linear uniform prior (covering the same range) for the number of MSPs per GC.}
    \label{fig:SBI-rea-NMSP-linear}
\end{figure}

The results of the inferred posterior distributions for Terzan 5, 47 Tuc and NGC 1851 (and the MSP gamma-ray luminosity function) are provided in Fig.~\ref{fig:SBI-rea-NMSP-linear}. We observe a quantitative shift in the obtained parameter values, consistent with the known parameter correlations. Qualitatively, the median prediction for the number of gamma-ray MSPs in each GC is larger by a factor of a few compared to the results assuming log-uniform priors. This shift is compensated by a lower pivot luminosity of the population and a larger width of the log-normal luminosity function. Clearly, if each GC hosts more sources at the same gamma-ray flux level, the average luminosity of an MSP in the respective population must be smaller. The enlarged width $\sigma_L$ reflects the necessity to produce MSPs still bright enough to explain the GCs with a single dominant MSP and those with significant ON- and OFF-phase emission. 

This behavior indicates that, given the present gamma-ray spectral information and the adopted forward model, the absolute MSP counts are only partially inferable and subject to the choice of priors: the likelihood exhibits an extended degeneracy in which an increase in $N_I$ can be compensated by a corresponding change in $L_0$ and the width of the luminosity function $\sigma_L$. However, since there is currently no robust, model-independent a priori knowledge of the number of gamma-ray-emitting MSPs hosted by an individual GC, we adopt a deliberately \emph{scale-free} prior for each $N_I$, i.e.~a prior that assigns comparable weight per decade in the number of MSPs. On a bounded interval, this corresponds to a log-uniform prior, which represents the most agnostic choice when the relevant scale of $N_I$ is uncertain over orders of magnitude. In contrast, a linear-uniform prior on the same interval places substantially more mass at large $N_I$ and therefore injects an implicit preference for high MSP counts. We therefore regard the log-uniform prior as the appropriate default for an agnostic estimate of $N_I$, while using the linear-uniform case as a prior-sensitivity test that quantifies the remaining multiplicity-luminosity degeneracy in the current data and modeling setup.

\section{Definitions of globular cluster datasets}
\label{app:gc-sets}

In this appendix, we list in Tab.~\ref{tab:gc-sets} the composition of each GC target dataset employed in the main text analysis, focusing on real LAT data. The associated GC SEDs have been extracted from 4FGL-DR4. The bright target dataset contains the SEDs of seven GCs that exhibit the highest total gamma-ray emission among the gamma-ray-bright GCs. In addition, it contains the GCs with known individual gamma-ray MSPs. 
Indeed, four GCs have gamma-ray pulsations detected and are associated with a known MSP. 
NGC 6341 and NGC 6652 were reported to host only a single, already detected MSP without significant OFF-phase emission~\citep{Zhang:2023gwr}. The same work also found that another two GCs, NGC 6624 and NGC 6626, host one detected MSP and feature significant OFF-phase gamma-ray emission, rendering their MSP population larger than 1. 
The remaining target datasets are to be understood as incremental subsets of each other, read from {left to right}. Therefore, we only list those GCs that are added to the selection at this level.

\begin{table*}[h]
    \centering
    \renewcommand{\arraystretch}{1.6}
    \begin{tabular}{c c c c}
        \toprule
        bright target data & intermediate target data & baseline target data &  extended target data   \\
        \midrule
        Terzan 5 & + & + & + \\
        47 Tuc & NGC 6316 & NGC 362 & M 10\\
        2MS-GC01 & $\omega$-Centauri & NGC 1851 & NGC 6569\\
        M 62 & NGC 6752 & NGC 5286 & M 19\\
        NGC 6388 & NGC 1904 & NGC 5904 & Whiting 1\\
        NGC 6441 & NGC 2808 & NGC 6139 & UKS 1\\
        NGC 6440 & NGC 6397 & NGC 6304 & M 69\\
        NGC 6652$(^*)$ & NGC 6541 & NGC 6402 & NGC 6539\\
        
        NGC 6341$(^*)$ & NGC 6656 & NGC 6528 & NGC 6522\\
        NGC 6624$(^\dagger)$ & NGC 6717 & NGC 6544 & \\
        NGC 6626$(^\dagger)$ & NGC 7078 & NGC 6712 & \\
         & NGC 6218 & Terzan 6 & \\
         & Palomar 6 & & \\
         & Terzan 1 & & \\
         & Terzan 2 & & \\
    \end{tabular}
    \caption{Compositions of each GC SED target dataset defined for the main text analysis using 4FGL-DR4 data products. \\ {$^*$: GCs hosting 1 detected MSP, but no significant OFF-phase emission regarding this MSP. $^\dagger$: GC hosting 1 detected MSP and significant OFF-phase emission with respect to it.}
    }
    \label{tab:gc-sets}
\end{table*}

In contrast, Tab.~\ref{tab:synthetic-SEDs} contains all GCs for which we have generated simulated SEDs following the prescription in Sec.~\ref{sec:generate-synthetic}. The table also contains the relevant parameters that describe these synthetic MSP populations regarding the number of sources per GC and the common MSP gamma-ray luminosity function parameters.

\begin{table*}
\centering
\renewcommand{\arraystretch}{1.6}
\begin{tabular}{l|l|l|c|c|c}
\toprule
Globular cluster & Alternate name & 4FGL-DR4 name & $L_{0}^{s}$ & $\sigma_{L}^{s}$ & $N_{X}^{s}$\tabularnewline
\midrule
NGC 6266 & M 62 & 4FGL J1701.2-3006 & \multirow{9}{*}{$3\times10^{33}\;\mathrm{erg}/\mathrm{s}$} & \multirow{9}{*}{$1.6$} & $12$\\
Terzan 2 & / & 4FGL J1727.6-3050  &  & & $7$\\
NGC 5139 & $\omega$-Centauri  & 4FGL J1326.6-4729 & &  & $5$\\
NGC 1904 & M 79 & 4FGL J0524.4-2413 &  &  & $3$\\
NGC 6752 & / & 4FGL J1910.8-6001 & &  & $2$\\
NGC 6218 & M 12 & 4FGL J1647.2-0154 & &  & $1$\\
NGC 6341 & M 92 & 4FGL J1716.8+4310 & &  & $1$\\
NGC 6652 & / & 4FGL J1835.7-3258 & &  & $1$\\
NGC 6624 & / & 4FGL J1823.5-3020 & &  & $1$ (ON-phase) / $7$ (OFF-phase)\\
\bottomrule
\end{tabular}
\caption{Parameter choices for all synthetic GC SEDs considered in this work.}
\label{tab:synthetic-SEDs}
\end{table*}

\end{document}